\documentclass[11pt]{article}
\usepackage{axodraw}
\usepackage{epsfig}
\usepackage{amsfonts}
\usepackage{amsmath}
\usepackage{bbm}
\usepackage{cite}
\allowdisplaybreaks[1]

\input pix.sty

\newcommand{\UI}{U^{ }_\rmii{\!$I$}}
\newcommand{\vev}{v_0}
\newcommand{\vT}{v_\rmii{$T$}}
\newcommand{\haat}[1]{{#1}}
\newcommand{\prop}{\Delta^{ }}
\newcommand{\dmdm}{DM$'$-DM~} % {$\overline{\mbox{DM}}$-DM~}
\newcommand{\nS}{n_\rmii{$S$}}
\newcommand{\nG}{n_\rmii{$G$}}
\newcommand{\mW}{m_\rmii{$W$}}
\newcommand{\mZ}{m_\rmii{$Z$}}
\newcommand{\mV}{m_\rmii{$V$}}
\newcommand{\mVt}{m_\rmii{$\widetilde V$}}
\newcommand{\mS}{m_\rmii{$S$}}
\newcommand{\mWt}{m_\rmii{$\widetilde W$}}
\newcommand{\mZt}{m_\rmii{$\widetilde Z$}}
\newcommand{\mQt}{m_\rmii{$\widetilde Q$}}
\newcommand{\mh}{m_\rmi{$h$}}
\newcommand{\mH}{m_\rmii{$H$}}
\newcommand{\mA}{m_\rmii{$A$}}

\newcommand{\Xs}{{X}^{ }}

\newcommand{\Ss}{{S}^{ }}

\renewcommand{\eq}{eq.~}
\renewcommand{\eqs}{eqs.~}

\renewcommand{\se}{sec.~}
\renewcommand{\ses}{secs.~}
\renewcommand{\fig}{fig.~}
\renewcommand{\figs}{figs.~}

\newcommand{\alphas}{\alpha_{\rm s}}

\newcommand{\rmO}{{\mathcal{O}}}

\newcommand{\CF}{C_\rmii{R}}

\def\lsi{\raise0.3ex\hbox{$<$\kern-0.75em\raise-1.1ex\hbox{$\sim$}}}
\def\gsi{\raise0.3ex\hbox{$>$\kern-0.75em\raise-1.1ex\hbox{$\sim$}}}
\newcommand{\lsim}{\mathop{\lsi}}
\newcommand{\gsim}{\mathop{\gsi}}

\newcommand{\nF}{n_\rmii{F}}
\newcommand{\nB}{n_\rmii{B}}
 \renewcommand{\nF}[1]{n_\rmii{F{#1}}}
 \renewcommand{\nB}[1]{n_\rmii{B{#1}}}

\newcommand{\rmii}[1]{{\mbox{\tiny\rm{#1}}}}

\newcommand{\re}{\mathop{\mbox{Re}}}
\newcommand{\im}{\mathop{\mbox{Im}}}

\newcommand{\Tint}[1]{{\hbox{$\sum$}\!\!\!\!\!\!\!\int\,}_{\!\!\!\!\raise-0.9ex\hbox{$\scriptstyle{#1}$}}}
\newcommand{\Tinti}[1]{{{\Sigma}\!\!\!\!\raise0.3ex\hbox{$\int$}_\rmii{${#1}$}}}

\newcommand{\bi}{\begin{itemize}}
\newcommand{\ei}{\end{itemize}}
\newcommand{\hide}[1]{ }

\newcommand{\deltabar}{\delta\!\!\!\raise0.7ex\hbox{--}\,}
\def\TAsc(#1,#2)(#3,#4,#5)%
{\SetWidth{2.0}\CArc(#1,#2)(#3,#4,#5)\SetWidth{1.0}}
\def\Lwidth{3}

\def\TAgl(#1,#2)(#3,#4,#5){\SetWidth{2.0}\PhotonArc(#1,#2)(#3,#4,#5){\Lwidth}%
{6.283 #3 mul 360 div #4 #5 sub #4 #5 sub mul sqrt mul Tdensity mul}%
\SetWidth{1.0}}
\def\TLgl(#1,#2)(#3,#4){\SetWidth{2.0}\Photon(#1,#2)(#3,#4){\Lwidth}
{#1 #3 sub #1 #3 sub mul #2 #4 sub #2 #4 sub mul add sqrt Tdensity mul}%
\SetWidth{1.0}}
\newcommand{\piC}[1]{\;\parbox[c]{120pt}{\begin{picture}(120,60)(0,0)
\SetWidth{1.0}\SetScale{1.2} #1 \end{picture}}\; }
\renewcommand{\picc}[1]{\;\parbox[c]{60pt}{\begin{picture}(60,30)(0,0)
\SetWidth{1.0}\SetScale{1.3} #1 \end{picture}}\; }
\def\Lwidth{1.3}

\newcommand{\piB}[1]{\;\parbox[c]{60pt}{\begin{picture}(60,40)(0,0)
\SetWidth{1.0}\SetScale{1.0} #1 \end{picture}}\;}
\newcommand{\piD}[1]{\;\parbox[c]{80pt}{\begin{picture}(80,40)(0,0)
\SetWidth{1.0}\SetScale{1.0} #1 \end{picture}}\;}
\def\Deca{\piB{%
 \SetWidth{1.0} 
 \Lsc(10,0)(30,0)
 \Lsc(10,20)(30,20)
 \Lsc(30,0)(30,20)
 \Photon(30,20)(55,20){1.0}{5}
 \Photon(30,0)(55,0){-1.0}{5}
 \Text(5,0)[c]{{$\scriptstyle H$}}
 \Text(5,20)[c]{{$\scriptstyle H$}}
 \Text(25,10)[c]{{$\scriptstyle A$}}
}}
\def\Decb{\piB{%
 \SetWidth{1.0} 
 \Lsc(10,0)(30,0)
 \Lsc(10,20)(30,20)
 \Lsc(30,0)(30,20)
 \Photon(30,20)(55,0){1.0}{7}
 \Photon(30,0)(40,8){-1.0}{3}
 \Photon(45,12)(55,20){-1.0}{3}
 \Text(5,0)[c]{{$\scriptstyle H$}}
 \Text(5,20)[c]{{$\scriptstyle H$}}
 \Text(25,10)[c]{{$\scriptstyle A$}}
}}
\def\Decc{\piB{%
 \SetWidth{1.0} 
 \Lsc(10,0)(25,10)
 \Lsc(10,20)(25,10)
 \Photon(25,10)(40,0){-1.0}{4}
 \Photon(25,10)(40,20){1.0}{4}
 \Text(5,0)[c]{{$\scriptstyle H$}}
 \Text(5,20)[c]{{$\scriptstyle H$}}
}}
\def\Decd{\piD{%
 \SetWidth{1.0} 
 \Lsc(10,0)(25,10)
 \Lsc(10,20)(25,10)
 \Lsc(25,10)(50,10)
 \COval(50.5,10)(1.5,1.5)(0){Black}{Black}
 \SetWidth{0.5}
 \Line(50,10)(60,20) 
 \Line(50,10)(63,13) 
 \Line(50,10)(63,7) 
 \Line(50,10)(60,0) 
 \Text(5,0)[c]{{$\scriptstyle H$}}
 \Text(5,20)[c]{{$\scriptstyle H$}}
 \Text(37,15)[c]{{$\scriptstyle h$}}
}}
\def\Dece{\piD{%
 \SetWidth{1.0} 
 \Lsc(10,0)(25,10)
 \Lsc(10,20)(25,10)
 \Photon(25,10)(50,10){1.0}{6}
 \COval(50.5,10)(1.5,1.5)(0){Black}{Black}
 \SetWidth{0.5}
 \Line(50,10)(60,20) 
 \Line(50,10)(63,13) 
 \Line(50,10)(63,7) 
 \Line(50,10)(60,0) 
 \Text(5,0)[c]{{$\scriptstyle A$}}
 \Text(5,20)[c]{{$\scriptstyle H$}}
 \Text(37,16)[c]{{$\scriptstyle Z$}}
}}
\def\Decf{\piD{%
 \SetWidth{1.0} 
 \Lsc(10,0)(25,10)
 \Lsc(10,20)(25,10)
 \Lsc(25,10)(50,10)
 \COval(50.5,10)(1.5,1.5)(0){Black}{Black}
 \SetWidth{0.5}
 \Line(50,10)(60,20) 
 \Line(50,10)(63,13) 
 \Line(50,10)(63,7) 
 \Line(50,10)(60,0) 
 \Text(5,0)[c]{{$\scriptstyle A$}}
 \Text(5,20)[c]{{$\scriptstyle H$}}
 \Text(37,15)[c]{{$\scriptstyle G$}}
}}
\def\NRa{\piC{%
 \SetWidth{1.0} 
 \Line(0,21.5)(20,21.5)%
 \Line(0,18.5)(20,18.5)%
 \CBox(18,18)(22,22){Black}{Yellow}
 \CArc(60,-10)(50,40,88)
 \CArc(60,-10)(50,92,140)
 \CArc(60,50)(50,220,268)
 \CArc(60,50)(50,272,320)
 \Photon(77,3)(77,37){1.5}{8}
 \COval(77,20)(4,4)(0){Black}{Gray}
 \Photon(43,3)(43,37){-1.5}{8}
 \COval(43,20)(4,4)(0){Black}{Gray}
 \Line(100,21.5)(120,21.5)%
 \Line(100,18.5)(120,18.5)%
 \CBox(98,18)(102,22){Black}{Yellow}
 \SetWidth{0.4}
 \Line(60,-5)(60,45) 
}}
\def\NRb{\piC{%
 \SetWidth{1.0} 
 \Line(0,21.5)(20,21.5)%
 \Line(0,18.5)(20,18.5)%
 \CBox(18,18)(22,22){Black}{Yellow}
 \CArc(60,-10)(50,40,61)
 \CArc(60,-10)(50,65,140)
 \CArc(60,50)(50,220,241)
 \CArc(60,50)(50,245,320)
 \Photon(60,0)(60,10){1.5}{2.5}
 \Photon(60,30)(60,40){1.5}{2.5}
 \Photon(87,8)(87,32){1.5}{6}
 \COval(87,20)(4,4)(0){Black}{Gray}
 \Photon(33,8)(33,32){-1.5}{6}
 \COval(33,20)(4,4)(0){Black}{Gray}
 \Line(100,21.5)(120,21.5)%
 \Line(100,18.5)(120,18.5)%
 \CBox(98,18)(102,22){Black}{Yellow}
 \SetWidth{0.4}
 \Line(21,-5)(99,45) 
 \Line(60,10)(66,13)%
 \Line(60,10)(54,13)%
 \Line(60,30)(66,27)%
 \Line(60,30)(54,27)%
}}
\def\NRZ{\piC{%
 \SetScale{1.0}
 \SetWidth{1.0} 
 \Line(20,45)(45,40)%
 \Line(45,40)(75,40)%
 \Line(75,40)(100,35)%
 \Line(20,-5)(45,0)%
 \Line(45,0)(75,0)%
 \Line(75,0)(100,5)%
 \Photon(75,0)(75,40){1.5}{8}
 \Photon(45,0)(45,40){-1.5}{8}
 \Text(35,20)[c]{{$\scriptstyle Z$}}
 \Text(85,20)[c]{{$\scriptstyle Z$}}
 \Text(10,-5)[c]{{$\scriptstyle {\rm{DM}}$}}
 \Text(60,-8)[c]{{$\scriptstyle {\rm{DM}'}$}}
 \Text(110,5)[c]{{$\scriptstyle {\rm{DM}}$}}
 \Text(10,45)[c]{{$\scriptstyle {\rm{DM}}$}}
 \Text(60,48)[c]{{$\scriptstyle {\rm{DM}'}$}}
 \Text(110,35)[c]{{$\scriptstyle {\rm{DM}}$}}
}}
\def\NRZp{\piC{%
 \SetScale{1.0}
 \SetWidth{1.0} 
 \Line(20,40)(100,40)%
 \Line(20,0)(100,0)%
 \Photon(75,0)(75,40){1.5}{8}
 \Photon(45,0)(45,40){-1.5}{8}
 \Text(35,20)[c]{{$\scriptstyle Z$}}
 \Text(85,20)[c]{{$\scriptstyle Z$}}
 \Text(10,0)[c]{{$\scriptstyle {\rm{DM}}$}}
 \Text(60,-8)[c]{{$\scriptstyle {\rm{DM}'}$}}
 \Text(110,0)[c]{{$\scriptstyle {\rm{DM}}$}}
 \Text(8,40)[c]{{$\scriptstyle {\rm{DM}'}$}}
 \Text(60,48)[c]{{$\scriptstyle {\rm{DM}}$}}
 \Text(112,40)[c]{{$\scriptstyle {\rm{DM}'}$}}
}}
\def\NRW{\piC{%
 \SetScale{1.0}
 \SetWidth{1.0} 
 \Line(20,45)(45,40)%
 \Line(45,40)(75,40)%
 \Line(75,40)(100,35)%
 \Line(20,-5)(45,0)%
 \Line(45,0)(75,0)%
 \Line(75,0)(100,5)%
 \Photon(75,0)(75,40){1.5}{8}
 \Photon(45,0)(45,40){-1.5}{8}
 \Text(32,20)[c]{{$\scriptstyle W^+$}}
 \Text(88,20)[c]{{$\scriptstyle W^-$}}
 \Text(10,-5)[c]{{$\scriptstyle {\rm{DM}}$}}
 \Text(110,5)[c]{{$\scriptstyle {\rm{DM}}$}}
 \Text(60,-8)[c]{{$\scriptstyle {\rm{X}}$}}
 \Text(10,45)[c]{{$\scriptstyle {\rm{DM}}$}}
 \Text(110,35)[c]{{$\scriptstyle {\rm{DM}}$}}
 \Text(60,48)[c]{{$\scriptstyle {\rm{X}}$}}
}}
\def\loss{\piD{%
 \SetWidth{1.0}
 \Photon(20,20)(50,20){1.5}{4} 
 \Line(40,30)(50,20)%
 \Line(50,20)(60,11)%
 \Text(8,20)[c]{{$\scriptstyle (\omega,\vec{k})$}}
 \Text(35,37)[c]{{$\scriptstyle(\epsilon^{ }_p,\vec{p})$}}
 \Text(62,5)[c]{{$\scriptstyle (\epsilon^{ }_{p+k},\vec{p+k})$}}
}}
\def\gain{\piD{%
 \SetWidth{1.0}
 \Photon(50,20)(20,20){1.5}{4} 
 \Line(30,30)(20,20)%
 \Line(20,20)(10,11)%
 \Text(62,20)[c]{{$\scriptstyle (\omega,\vec{k})$}}
 \Text(35,37)[c]{{$\scriptstyle(\epsilon^{ }_p,\vec{p})$}}
 \Text(8,5)[c]{{$\scriptstyle (\epsilon^{ }_{p+k},\vec{p+k})$}}
}}
\makeatletter \@addtoreset{equation}{section} \makeatother
\renewcommand{\theequation}{\arabic{section}.\arabic{equation}}
\makeatletter
\renewcommand\section{\@startsection {section}{1}{\z@}%
                                   {-5.5ex \@plus -1ex \@minus -.2ex}% bfr-
                                   {2.3ex \@plus.2ex}%
                                   {\normalfont\large\bfseries}}
\renewcommand\subsection{\@startsection{subsection}{2}{\z@}%
                                     {-3.25ex\@plus -1ex \@minus -.2ex}%
                                     {1.5ex \@plus .2ex}%
                                     {\normalfont\normalsize\bfseries}}
\renewcommand\thesection {\@arabic\c@section}
\renewcommand\thesubsection   {\thesection.\@arabic\c@subsection}
\renewcommand{\@seccntformat}[1]{%
\csname the#1\endcsname.\hspace{1.0em}}
\makeatother
\begin{document}

\flushbottom

\begin{titlepage}

\begin{flushright}
% DRAFT M.L. \\ 
% arXiv:1609.00474\\ 
January 2017
\vspace*{1cm}
\end{flushright} 
\begin{centering}

\vfill

{\Large{\bf
  On thermal corrections to near-threshold annihilation 
}} 

\vspace{0.8cm}

Seyong Kim$^{\rm a}$ and %\footnote{skim@sejong.ac.kr}
M.~Laine$^{\rm b}$ %\footnote{laine@itp.unibe.ch}

\vspace{0.8cm}

$^\rmi{a}$%
{\em
Department of Physics, Sejong University, 
Gunja-Dong 98, Seoul 143-747, South Korea\\}

\vspace*{0.3cm}

$^\rmi{b}$%
{\em
AEC, Institute for Theoretical Physics, 
University of Bern, \\ 
Sidlerstrasse 5, CH-3012 Bern, Switzerland\\} 

\vspace*{0.8cm}

\mbox{\bf Abstract}

\end{centering}

\vspace*{0.3cm}
 
\noindent
We consider non-relativistic ``dark'' particles
interacting through gauge boson exchange.  At finite temperature, gauge
exchange is modified in many ways: virtual corrections
lead to Debye screening; real corrections amount to frequent
scatterings of the heavy particles on light plasma constituents;
mixing angles change. In a certain temperature and energy range, 
these effects are of order unity. Taking them into account in a
resummed form, we estimate the near-threshold spectrum of 
kinetically equilibrated annihilating TeV scale particles.  
Weakly bound states are shown to ``melt'' below
freeze-out, whereas with attractive strong interactions, relevant e.g.\ for
gluinos, bound states boost the annihilation rate by a factor $4...80$ 
with respect to the Sommerfeld estimate, thereby perhaps helping to
avoid overclosure of the universe. Modestly non-degenerate dark sector 
masses and a way to combine the contributions of channels 
with different gauge and spin structures are also discussed. 

\vfill

%\noindent
%PACS numbers: 
%11.10.Wx, %        Finite temperature field theory
%11.15.Ha, %        Lattice gauge theory 
%12.38.Bx, %        Perturbative calculations in QCD
%12.38.Mh, %        Quark--gluon plasma
%14.40.Nd, %        Bottom mesons
%\\
%Keywords:
 
%% \vspace*{1cm}
%%   
%% \noindent
%% June 2016

\vfill

\end{titlepage}

%%%%%%%%%%%%%%%%%%%%%%%%%%% SECTION %%%%%%%%%%%%%%%%%%%%%%%%%%%%%%%%%%%%%%
%
\section{Introduction}

The possibility that stable or long-lived massive neutral particles
could be responsible for dark matter,  continues to motivate
a versatile program of direct and indirect searches and
collider experiments. The cosmological
abundance of such particles is determined by a ``freeze-out'' process, 
taking place when the annihilation rate decreases below the Hubble 
rate. For particles of mass $M$, the freeze-out temperature is generically 
of order $T \sim M/25 ... M/20$.\footnote{%
 This follows from $H \sim n \langle \sigma v \rangle$, 
 i.e.\ 
 $\frac{T^2}{m_\rmii{Pl}} \sim \bigl( \frac{MT}{2\pi} \bigr)^{3/2} e^{-M/T}
 \frac{\alpha^2}{M^2}$, where $\alpha$ is some fine-structure constant. 
 } 
In this regime the particles are kinetically
equilibrated and non-relativistic. Therefore they move slowly and 
have time to experience repeated interactions 
(cf.~e.g.~refs.~\cite{old1,old2,hisano,sfeldx,feng}).  

It is conceivable that repeated soft 
interactions could modify the nature of 
the annihilation process. For instance, 
it has been appreciated in recent years that in certain models
there are attractive interactions between dark matter particles, or 
between particles co-annihilating with dark matter particles,  
which could lead 
to bound-state formation even in weakly interacting cases 
(cf.~e.g.~refs.~\cite{wimpo1,wimpo2,wimpo3,wimpo4}).  
A number of studies 
(cf.~e.g.~refs.~\cite{old3,old32,old35,old4,old5,old52,old53}) 
have included bound states in a freeze-out analysis, 
notably by adding an on-shell bound state phase space distribution
as an independent degree of freedom in a set of Boltzmann equations.
A thereby increased annihilation rate might represent 
a phenomenologically welcome development, given that LHC 
searches have pushed up the dark matter mass scale, which could 
lead to the weakly interacting
dark matter energy density overclosing the universe. 
 
Treating bound states precisely is a non-trivial task, 
and furthermore quite sensitive to thermal effects~\cite{ms}. 
In ref.~\cite{4quark_lattice}, basic formulae for the inclusion 
of bound states
on the perturbative and non-perturbative levels were derived, 
working within the framework of non-relativistic effective 
field theories~\cite{nrqcd,bodwin}.
The formalism was also applied to a particular model, 
QCD at $T \gsim 150$~MeV.  
Both a perturbative and a lattice study found an enhancement 
of the singlet channel annihilation rate of bottom quarks
by up to two orders of magnitude with respect
to a previous estimate~\cite{pert}, which was based on a thermally 
averaged ``Sommerfeld factor''~\cite{asommerfeld,landau3,fadin,new}, 
correcting the annihilation rate of free scatterers. 

The purpose of the present paper is to apply the perturbative
side of the approach
of ref.~\cite{4quark_lattice} to simple examples in cosmology. 
In particular, we show that thermal corrections to the near-threshold
spectrum (or the differential annihilation rate)
are of order unity in a temperature range (\eq\nr{TvsM})
which may coincide with that of the freeze-out process. 
The effect on the total annihilation rate is in general small in weakly
coupled systems, whereas in strongly coupled systems near-threshold
annihilation can dominate the total rate. 

To put the physics in a wider context, we note in passing that thermal
corrections to annihilation phenomena have been addressed in great detail 
in the context of nuclear reactions 
in astrophysical plasmas~(cf.\ ref.~\cite{lsb} for a review). Those processes
resemble the present ones in the sense that the energy released
is large compared with thermal scales, and that
the annihilation process is accurately captured by 
effective four-particle operators. Of course, 
there is the qualitative difference that the Coulomb interaction
between the non-relativistic ionized nuclei is repulsive, 
so that no bound states can form. 

The paper is organized as follows. 
We start by discussing 
the energy and temperature scales 
relevant for non-relativistic annihilation in 
\se\ref{se:physics}.
A thermally averaged $s$-wave annihilation rate and a corresponding 
spectral function are defined 
in \se\ref{se:theory}, where we also recall 
how these can be computed in resummed perturbation theory, 
accounting for collective plasma phenomena which yield
the dominant thermal corrections. 
Sec.~\ref{se:Z} contains an application of the formalism 
to the case of dark particles bound together 
by Standard Model $Z$ exchange, \se\ref{se:Zp} to 
light dark $Z'$ exchange, and 
\se\ref{se:gluon} to gluon exchange. 
In \se\ref{se:nondeg} we discuss how the situation changes if the 
dark particles are modestly non-degenerate in mass, and in \se\ref{se:decomp}
how different annihilation channels can be combined. 
Conclusions and an outlook are offered in \se\ref{se:concl}. 
In two appendices the transverse parts of thermal $Z$ and $Z'$ 
self-energies are computed at 1-loop order in a general 
$R^{ }_\xi$ gauge, demonstrating the gauge independence of 
the structures that affect our thermal considerations. 

%%%%%%%%%%%%%%%%%%%%%%%%%%% SECTION %%%%%%%%%%%%%%%%%%%%%%%%%%%%%%%%%%%%%%
%
\section{Physics background: scales in a thermal medium}
\la{se:physics}

In order to introduce the various phenomena that play a role, we start 
by defining a number of energy and momentum scales affecting the dynamics. 
Subsequently examples of how the scales interfere with each other
are outlined. 

\paragraph{\em Non-relativistic energy and momentum.} 
Kinetically equilibrated non-relativistic particles 
of mass $M$ at a temperature $T$ move with
an average velocity $v\sim (T/M)^{1/2} \ll 1$ and 
have a kinetic energy $E_\rmi{kin} \sim Mv^2 \sim T \ll M$. 
If the particles interact through Coulomb-like exchange,  the 
associated potential energy is $E_\rmi{pot} \sim \alpha/r \sim M v  \alpha$, 
where we expressed the typical distance between the annihilating particles,
$r$, through 
the uncertainty relation as the inverse relative momentum, $r\sim 1/(M v)$.
For $v > \alpha$ the potential energy is small compared with the 
kinetic energy, but for $v\sim\alpha$ the two are of the same order, 
leading to Sommerfeld corrections of order unity. 
If the particles happen to form a bound state, they can no longer
move freely, but we can still speak of an average velocity associated
with the bound motion. In this case $E_\rmi{kin} \sim E_\rmi{pot}$
by definition, 
so that $v \sim \alpha$. Thereby the binding energy associated with
bound states is $\Delta E \sim M\alpha^2$.

\paragraph{\em Thermal widths.}
An interacting particle gets constantly kicked by  
scatterings with plasma constituents. The scatterings imply that the 
particle has a finite ``width'', or interaction rate. This does not 
mean that the particle would decay, but that it can change its phase or 
colour or momentum or go into an excited state. 
Parametrically, for a single heavy particle, 
the width is $\Gamma_\rmi{int} \sim \alpha T$~\cite{ht1}. 
(No momentum transfer is involved in these scatterings; if we wish to 
adjust momenta, the relevant concept is the kinetic equilibration rate, 
which scales as $\Gamma_\rmi{kin} \sim \alpha^2 T^2/M$~\cite{gamma_kin}.)  
If we consider a pair of heavy particles attracting each other through
gauge exchange, then the interaction rate is  
smaller than $2 \Gamma_\rmi{int}$, because 
close to each other the particles 
would form a gauge singlet, which does not feel
gauge interactions. In fact the width has a ``dipole'' shape
at small separations, 
$\Gamma \sim \alpha^2 T^3 r^2$~\cite{static}. 
Inserting $r\sim 1/(M v)$, this leads to 
$\Gamma \sim \alpha^2 T^2 / M$ for scattering states, 
and $\Gamma \sim T^3 / M^2$ for bound states. In the case
of scattering states, with $E \sim T$, the width 
plays a subleading role, whereas for bound states 
the issue is more subtle and is discussed below. 

\paragraph{\em Thermal masses.}
Apart from thermal widths, 
thermal effects also lead to ``virtual corrections'', notably thermal masses.
For electric fields responsible for the Coulomb-like exchange,
the thermal mass is known as a Debye mass and is of order 
$m_\rmi{th} \sim\alpha^{1/2}T$. This defines
the distance scale at which 
gauge exchange varies; for instance, the width defined above is of the form
$\Gamma \sim \Gamma_\rmi{int}\, \Phi(m_\rmi{th} r)$, 
with $\Phi(x) \sim x^2$ for $x \ll 1$ and $\Phi(x) = 2$ for $x\gg 1$.

The heavy particles also experience 
thermal mass shifts. An unresummed perturbative computation yields 
$\delta M_\rmi{th} \sim \alpha T^2/M$~\cite{dhr},\footnote{%
 In the dark matter context
 mass corrections of this type were considered and shown 
 to be small in ref.~\cite{tw}. 
 } 
however the Debye screening
of the electric field leads to a correction with a different structure
and an opposite sign (cf.\ \eq\nr{selfE}), 
\be
 \delta M_\rmi{rest,th} = - \fr12 \alpha m_\rmi{th} \sim - \alpha^{3/2} T
 \;. \la{Salpeter}
\ee  
If $T < \alpha^{1/2} M$, as is the case in our
considerations, the latter correction dominates. 
This is the case in general: unresummed perturbation theory leads to
power-suppressed thermal corrections, but collective plasma phenomena
yield larger effects (a nice discussion can be found in sec.~6 of 
ref.~\cite{chesler}). 
In the context of nuclear rates 
\eq\nr{Salpeter} amounts to 
a ``Salpeter correction'' (cf.\ ref.~\cite{lsb} for a review), 
which increases the annihilation 
rate by a factor $\exp(-2 \delta M_\rmi{rest,th}/T) 
= \exp(\alpha m_\rmi{th} / T)$.
This is a correction of $\rmO(\alpha^{3/2})$ to the total
rate, but an $\rmO(1)$ effect close to the threshold, given that its 
location gets shifted. 

\paragraph{\em When does the Sommerfeld effect play a role
for annihilation?}
Consider scattering states with $v\sim (T/M)^{1/2}$. 
As discussed above, the potential
energy from a Coulomb exchange is of the same order
as the kinetic energy for $v \sim\alpha$. 
Therefore, from $(T/M)^{1/2} \sim \alpha$,  
we find that the Sommerfeld effect is of order unity for 
\be
 T \sim \alpha^2 M
 \;. \la{TvsM2}
\ee 
In contrast, 
in the range $T \gsim \alpha M$ to be defined in \eq\nr{TvsM}, 
where $v\gsim \alpha^{1/2}$, 
$E^{ }_\rmi{pot}$ only represents a subset of higher-order corrections. 
It may be noted that the momenta exchanged by scattering states 
are large compared with Debye masses, 
$M v\sim (MT)^{1/2} \gg m_\rmi{th} \sim \alpha^{1/2} T$, 
and the kinetic energy of the annihilating pair is large compared with its
thermal width, $T \gg \Gamma \sim \alpha^2 T^2/M$.
Therefore thermal effects can be omitted from
Sommerfeld considerations at leading order 
in $\alpha$~\cite{pert}. However we do 
expect an effect of $\rmO(\alpha^{3/2})$ as shown by 
\eq\nr{Salpeter}, and return to a discussion of 
the magnitude of thermal effects below \eq\nr{regime}. 

\paragraph{\em When do bound states exist in a thermal medium?}
Consider an {\em attractive} Coulomb-like exchange, $V(r) = -\alpha/r$.
A conservative estimate asserts that bound states ``melt'' when the thermal 
screening length (inverse of $m_\rmi{th}$)
has become shorter than the Bohr radius~\cite{ms}, 
$1/(\alpha^{1/2} T) \lsim 1/(\alpha M)$, i.e.\ $T \gsim \alpha^{1/2}M$.  
A more stringent estimate is obtained by requiring that the thermal width
exceeds the binding energy:
$\Gamma \sim T^3 / M^2 \,\gsim\, 
\Delta E \sim \alpha^2 M$, 
i.e.\ $T \gsim \alpha^{2/3}M$~\cite{soto,review,wu}.
However it is difficult to fix the prefactor of this estimate, 
and therefore to decide whether in the case of 
weak interactions, with $\alpha \sim 10^{-2}$, bound states 
can persist up to the temperatures
$T \sim M/25 ... M/20$ that are relevant for the freeze-out analysis. 
A numerical investigation 
is carried out for various models in \ses\ref{se:Z} and \ref{se:Zp}, 
cf.\ figs.~\ref{fig:Z0} and \ref{fig:Z0p}. 
For simple power counting, 
we consider the regime 
\be 
 T \gsim \alpha M
 \la{TvsM}
\ee 
for this purpose, 
in which case bound states exist at $T\sim \alpha M$ and then melt
at $T \gg \alpha M$, 
i.e.\ thermal corrections are of order unity. In practice
the gauge exchange is typically Yukawa screened, but for simplicity
we use the Coulombic estimate in the following.

\paragraph{\em When do bound states play a role
for annihilation?}
If bound states exist, they have a binding energy 
$\Delta E \sim M\alpha^2$. The Boltzmann weight is then 
boosted by a factor $\exp(\Delta E/T)$, implying that
bound states have an effect of order unity for $T \sim \alpha^2 M$, 
just like the Sommerfeld effect in \eq\nr{TvsM2}. 
In the regime of \eq\nr{TvsM}, in contrast,     
bound-state contributions amount to higher-order
corrections to the total annihilation rate, because 
$\Delta E/T \lsim \alpha$. A strongly interacting case 
in which bound states do dominate the total annihilation rate
is discussed in \se\ref{se:gluon}, cf.\ \fig\ref{fig:gluon}.

%%%%%%%%%%%%%%%%%%%%%%%%%%% SECTION %%%%%%%%%%%%%%%%%%%%%%%%%%%%%%%%%%%%%%
%
\section{Theoretical framework}
\la{se:theory}

In order to address the phenomena outlined above, 
we formulate a specific theoretical framework. 
We start by defining a thermally averaged annihilation 
rate in the non-relativistic regime (\se\ref{ss:decay}); show
how the energy scales contributing to the thermal annihilations can be 
resolved through a spectral function (\se\ref{ss:rho}); recall
how the spectral function can be determined (beyond 
strict perturbation theory) through the solution
of an inhomogeneous Schr\"odinger equation (\se\ref{ss:Sch}); 
and discuss how the ``static potential'' appearing in the 
Schr\"odinger equation can be computed within a thermal 
plasma (\se\ref{ss:prop}). 
We refer to the heavy particles as
DM and DM$'$, even though the two species can also be the same
(cf.\ \fig\ref{fig:exchange}). 

%%%%%%%%%%%%%%%%%%%%%%%%%% FIGURE %%%%%%%%%%%%%%%%%%%%%%%%%%%%%%%%%%%%%%%%%
%
\begin{figure}[t]
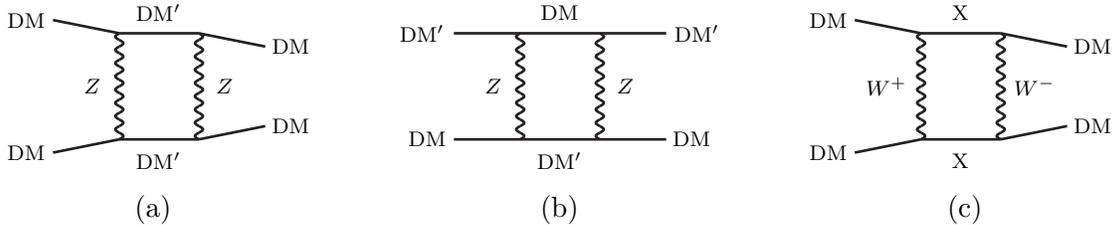


\hspace*{6mm}%
\begin{minipage}[c]{14.2cm}
\begin{eqnarray*}
&& 
 \hspace*{-1.2cm}
 \NRZ
 \hspace*{0.9cm}
 \NRZp
 \hspace*{0.9cm}
 \NRW
\\[6mm]
&& \hspace*{7mm}
   \mbox{(a)}
   \hspace*{4.9cm}
   \mbox{(b)}
   \hspace*{4.9cm}
   \mbox{(c)}
\end{eqnarray*}
\end{minipage}

\vspace*{1mm}

\caption[a]{\small 
 Examples of processes with static gauge exchange. 
 Left: with neutral gauge exchange, the particle identity 
 remains the same in the case of fermions (DM$'$ = DM) 
 but changes for scalars (DM$' \neq$  DM). The diagram illustrates
 the kinematics for the case $m^{ }_\rmii{DM$'$} > m^{ }_\rmii{DM}$. 
 Middle: another possibility for gauge exchange. A further one, relevant
 for certain models, can be obtained by exchanging DM and DM$'$
 in the intermediate stage.  
 Right: with charged gauge exchange, particle identity  
 necessarily changes (X $\neq$ DM). If $m^{ }_\rmii{X} \gg m^{ }_\rmii{DM}$,  
 only $Z$ exchange needs to be considered. 
 }
\la{fig:exchange}
\end{figure}
%
%%%%%%%%%%%%%%%%%%%%%%%%%%%%%%%%%%%%%%%%%%%%%%%%%%%%%%%%%%%%%%%%%%%%%%%%%%

%%%%%%%%%%%%%%%%%%%%%%%%%%% SUBSECTION %%%%%%%%%%%%%%%%%%%%%%%%%%%%%%%%%%%
%
\subsection{Thermally averaged annihilation rate and equilibrium 
number density}
\la{ss:decay}

Let $\eta\theta$ stand for a local operator 
which annihilates a \dmdm pair. Eigenstates of the Hamiltonian
containing a \dmdm pair, either a bound or a scattering state, 
are denoted by $|m\rangle$ and have energies $E^{ }_m \sim 2 M$.\footnote{%
 For simplicity of notation we assume the system to be placed in a large
 periodic box, so that the spectrum of scattering states is 
 discrete, however in the end the thermodynamic limit is taken. 
 } 
Within non-relativistic theories~\cite{nrqcd}, 
inclusive $s$-wave \dmdm annihilations 
can be described by local four-particle operators of the type 
$
 \mathcal{O} = 
 i c^{ }_1\alpha^2\, \theta^\dagger \eta^\dagger\, \eta\theta / M^2
$~\cite{bodwin}, 
where $\alpha$ is a fine structure constant evaluated at
a hard renormalization scale $\sim 2 M$ and 
$c^{ }_1$ is a group-theoretic coefficient.  Within a thermal 
medium, the annihilations mediated by this operator define 
a ``chemical equilibration rate'', $\Gamma_\rmi{chem}$, implying that the dark 
matter density $n$ evolves as  
\be
 (\partial_t + 3 H )\, n = - \Gamma_\rmi{chem}(n- n_\rmi{eq}) + 
 \rmO(n - n_\rmi{eq})^2
 \;. \la{linear_response}
\ee
Here $H$ is the Hubble rate and 
$n^{ }_\rmi{eq}$ is the DM equilibrium number density.
Through a linear response analysis, $\Gamma_\rmi{chem}$ can be 
related to an equilibrium 2-point correlator and then expressed
as a ``transport coefficient''~\cite{chemical}. Within the NRQCD
framework the transport coefficient turns out to be 
proportional to the intuitively transparent 
thermal expectation value~\cite{4quark_lattice} 
\be
 \gamma \;\equiv\; \frac{1}{\mathcal{Z}}
 \sum_{m} e^{-E^{ }_m / T } 
 \langle m | \theta^\dagger\eta^\dagger\, \eta\theta | m \rangle
 \;, \la{def_Gamma}
\ee
as
$
 \Gamma^{ }_\rmi{chem} \approx
 8 c^{ }_1 \alpha^2 \gamma / (M^2 n_\rmi{eq})
$.  Linearizing
a dark matter Boltzmann equation~\cite{clas1,clas2}\footnote{%
  For simplicity we consider a single DM species here
  (with DM$'$ equivalent or antiparticle to DM), 
  with $N$ internal degrees of freedom; systems with multiple
  non-degenerate species are addressed
  in \ses\ref{se:gluon} and \ref{se:nondeg}.}, {\it viz.}
\be
 (\partial_t + 3 H )\, n \simeq - \langle \sigma v \rangle\,
 (n^2 - n_\rmi{eq}^2) 
 \;, \la{Boltzmann}
\ee
we can identify 
$\langle \sigma v \rangle = \Gamma_\rmi{chem} /( 2 \, n_\rmi{eq})$  
and therefore get
$
 \langle \sigma v \rangle \approx
 4 c^{ }_1 \alpha^2 \gamma / (M^2 n_\rmi{eq}^2)
$.
If the DM and DM$'$ particles have $N$ internal degrees of freedom, then
in the free limit $\gamma \to \gamma^{ }_\rmi{free} = n^2_\rmi{eq} / (4 N)$, 
cf.\ the discussion 
below \eq\nr{Gamma_free}. A further useful quantity, 
closely related to $\langle \sigma v \rangle$, is a ``thermally averaged
Sommerfeld factor'', characterizing the strength of interactions:
$\bar{S}^{ }_1 \equiv \gamma / \gamma^{ }_\rmi{free} 
= 4 N \gamma / n^2_\rmi{eq}$.
Thereby 
$
 \langle \sigma v \rangle \approx 
 c^{ }_1 \alpha^2 \bar{S}^{ }_1 / (M^2 N)
$.

In order to solve \eq\nr{linear_response} or \nr{Boltzmann}, 
we need to know the value of $n^{ }_\rmi{eq}$, 
and if we discuss radiative corrections 
to $\Gamma^{ }_\rmi{chem}$ or $\langle \sigma v \rangle$, 
we should also discuss those to $n^{ }_\rmi{eq}$. 
In order to compute such corrections, $n^{ }_\rmi{eq}$ has to be 
properly defined. 
As suggested in ref.~\cite{chemical}, it is physically meaningful  
to define $n \equiv e/M$, where $e$ is the energy
density carried by the dark matter particles. However,  
if dark matter is made of ``particles'' and ``antiparticles'', a simpler
definition is provided by the susceptibility related to a 
conserved Noether charge: 
$
 n^{ }_\rmi{eq} \equiv \chi^{ }_\rmi{f} \equiv 
 \frac{1}{V} \langle Q^2 \rangle
$
where $V$ is the volume and, 
for fermions, $Q = \int_{\vec{x}}\bar\psi\gamma_0 \psi$.
Indeed, evaluating this in the energy eigenbasis, we get
$
 n^{ }_\rmi{eq} = \frac{2}{\mathcal{Z}\, V} 
 \sum_{\vec{p}} e^{-\mathcal{E}^{ }_p /T} 
 + \rmO(e^{-2 M/T})
$, 
where $\mathcal{E}^{ }_p$ are the energy 
eigenvalues for states with a single heavy
particle and the factor 2 accounts for the antiparticles. Going
over to infinite volume and carrying out a resummed 
next-to-leading order (NLO) computation, we find
\be
 n^{ }_\rmi{eq} = 2 N 
 \int_{\vec{p}} e^{-\mathcal{E}^{ }_p/T}
 \, \biggl[
    1 - \frac{g^2 T^2 \CF}{12 p^2} + 
   \frac{g^2 m^{ }_\rmi{th} \CF}{8\pi T} 
    +\rmO(g^4,e^{-M/T})
 \biggr]
 \;, \quad
 \mathcal{E}^{ }_p \equiv \sqrt{p^2 + M^2} 
 \;, 
 \la{neq}
\ee
where $\CF$ is the quadratic Casimir of the gauge representation, 
$M$ corresponds technically to a pole mass, and $m^{ }_\rmi{th}$
is the Debye mass defined around \eq\nr{Salpeter}. Through partial 
integrations it can be shown that the first correction
amounts to the thermal mass of refs.~\cite{dhr,tw}, 
$M^2 \to M^2_\rmi{th} \equiv M^2 + \Delta M^2_\rmi{th}$, 
with 
$
 \Delta M^2_\rmi{th} = g^2 T^2 \CF/6
$. 
Both the ``rest'' and ``kinetic'' masses get corrected
by the same amount. The second correction amounts to the Salpeter
term in \eq\nr{Salpeter}, which only affects the rest mass. 
The latter term dominates if $T \ll g M$, because the 
average momentum is $p^2 \simeq M T$. 
This formally dominant contribution was omitted in
the unresummed computations of refs.~\cite{dhr,tw}, and can only be found
by properly incorporating Debye screening in the gauge 
field propagator. 
If we ``resum'' both 
corrections into the exponent and denote 
$\alpha \equiv g^2 \CF / (4\pi)$, then 
\be
 n^{ }_\rmi{eq} \approx 2 N \, 
 \Bigl( \frac{M^{ }_\rmi{th}T}{2\pi} \Bigr)^{3/2} 
 \exp\Bigl( -\frac{M^{ }_\rmi{th}}{T} + \frac{\alpha m_\rmii{th}}{ 2T} \Bigr)
 \;. \la{neq_final} 
\ee
We note that in $\gamma/n_\rmi{eq}^2$, which appears 
in $\langle \sigma v \rangle$ and $\bar{S}^{ }_1$ 
defined below \eq\nr{Boltzmann}, the latter term in the exponent
cancels against the Salpeter correction discussed below \eq\nr{Salpeter}. 

%%%%%%%%%%%%%%%%%%%%%%%%%%% SUBSECTION %%%%%%%%%%%%%%%%%%%%%%%%%%%%%%%%%%%
%
\subsection{Definition of a spectral function}
\la{ss:rho}

We now wish to resolve the total rate in \eq\nr{def_Gamma}
into a spectral representation, which tells
which kind of states are responsible for the annihilations. For this purpose
we first define a Wightman function,  
\be
 \Pi^{ }_{<}(\omega) \; \equiv \; 
 \int_{-\infty}^{\infty} \! {\rm d}t \, e^{i \omega t}
 \, \bigl\langle
 (\theta^\dagger\eta^\dagger)(0,\vec{0}) \, (\eta\theta)(t,\vec{0})
 \bigr\rangle^{ }_T
 \;, \la{Pi_small}
\ee
where $\langle ... \rangle^{ }_T$ refers to a thermal expectation value
and $\omega$ corresponds to the energy
released in the hard process. 
Clearly, 
the full rate in \eq\nr{def_Gamma}
is obtained from the integral over all possibilities, 
\be
 \gamma\; =\; \int_{-\infty}^{\infty} \! \frac{{\rm d}\omega}{2\pi}
 \, \Pi^{ }_{<}(\omega) 
 \;. \la{Pi_Gamma}
\ee

Now, for a better physical understanding, we re-express
\eq\nr{Pi_Gamma} in terms of a central underlying object, 
the {\em spectral function}. In operator language it is defined as 
\be
  \rho(\omega,\vec{k}) \; \equiv \; 
  \int_{-\infty}^{\infty} \! {\rm d}t \,
  \int^{ }_{\vec{r}} \, e^{i (\omega t - \vec{k}\cdot\vec{r})}
  \, \Bigl\langle
  \fr12 \bigl[ (\eta\theta)(t,\vec{r}), \, 
  (\theta^\dagger\eta^\dagger)(0,\vec{0}) \bigr] 
  \Bigr\rangle^{ }_T
 \;. \la{rho_k}
\ee 
We refer to $k \equiv |\vec{k}|$
as the total momentum of the pair with respect to the heat bath. 
All other 2-point correlators can be expressed in terms of the spectral
function, in particular
$
 \Pi^{ }_{<}(\omega) = 2 \nB{}(\omega) \int_{\vec{k}} \rho(\omega,\vec{k}) 
$, 
where $\nB{}$ is the Bose distribution. Inserting this information
into \eq\nr{Pi_Gamma}, assuming $\pi T \ll M$, and noting
that there is spectral weight only at $\omega \gsim 2 M$,  we obtain
\be
 \gamma = \int_{2M - \Lambda}^{\infty} 
 \! \frac{{\rm d}\omega}{\pi} \, e^{-\omega/T} \, 
 \int_{\vec{k}} \rho(\omega,\vec{k}) 
 \; + \; \rmO\bigl(e^{-4M/T}\bigr)  
 \;, \quad
 \alpha^2 M \ll \Lambda \;\lsim\; M 
 \;. \la{Laplace}
\ee
The cutoff $\Lambda$ plays no practical role as long as it 
is $\gg \alpha^2 M$, given that the spectral function vanishes for
$0 \ll \omega \ll 2M$;\footnote{%
 To be precise, 
 at finite temperature the spectral function does not vanish 
 exactly in this regime but has a small tail~\cite{nlo}; 
 the corresponding contribution to $\gamma$
 is suppressed by $\alpha$ and powers of $T/M$. 
 } 
nevertheless we introduce it in order to restrict the 
average to a regime in which a non-relativistic treatment and the 
replacement of the Bose distribution through the Boltzmann distribution
are formally justified. 

We note that 
the spectral function is a nice object because it is of 
$\rmO(1)$ rather than exponentially suppressed; 
the exponential suppression
has been factored into \eq\nr{Laplace}.  
In the following we sometimes refer to $\rho$ as a
{\em differential annihilation rate}, with the 
understanding that $\rho$ is to be weighted by $e^{-\omega/T}$
to properly fill this role. 

A final ingredient for applying \eq\nr{Laplace} is to note that, 
as usual in a non-relativistic two-body problem, the dependence 
on the total momentum $\vec{k}$ factorizes from the internal dynamics. 
Therefore it is sufficient to compute $\rho(\omega,\vec{k})$ 
for $\vec{k} = \vec{0}$, and recall afterwards that for $\vec{k} \neq \vec{0}$
the center-of-mass energy is $2 M + k^2/(4 M)$ rather than $2M$, 
cf.\ \eq\nr{Eprime}. 

%%%%%%%%%%%%%%%%%%%%%%%%%%% SUBSECTION %%%%%%%%%%%%%%%%%%%%%%%%%%%%%%%%%%%
%
\subsection{Ways to determine the spectral function}
\la{ss:Sch}

According to \eq\nr{Laplace}, we need to determine the spectral
function in the range
\be 
 |\omega - 2 M| \; \lsim \; \pi T \; \ll \; M
 \;. \la{regime}
\ee 
This puts us deep in the non-relativistic regime. In principle, spectral
functions can be computed in strict perturbation theory both in vacuum
and including thermal corrections. Thermal corrections can be shown
to be infrared (IR) finite at NLO, 
power-suppressed, and numerically small~\cite{nlo,mb1}, 
like the thermal mass of refs.~\cite{dhr,tw} which 
emerges as a part of these corrections~\cite{nlo}. However,   
as discussed around \eq\nr{Salpeter} and \eq\nr{neq}, 
these power-suppressed thermal 
corrections are in general 
not the dominant ones in the regime of \eq\nr{TvsM}; 
thermal corrections exist which are only suppressed by the coupling, 
not by $T^2/M^2$.  
In order to incorporate the dominant corrections
close to threshold, both at $T=0$ and
at $T> 0$, a suitable resummed framework is needed. 

Before proceeding to the resummed framework, 
it is appropriate to stress that the total annihilation 
rate from \eq\nr{def_Gamma} can be related to a purely Euclidean 
(imaginary-time) correlator~\cite{4quark_lattice}. Systematic 
higher-order perturbative computations and lattice studies should probably
take the imaginary-time formulation as a starting point. 

A way to compute resummed 
thermal spectral functions in the non-relativistic 
regime has been suggested in refs.~\cite{resum,peskin}. 
The power counting
behind this framework has 
been discussed in great detail in ref.~\cite{jacopo}, 
and corresponds to \eq\nr{TvsM}.\footnote{%
 Technically, the resummed framework assumes that the vacuum energy
 scale $\sim 2M$ and the thermal scale $\sim \pi T$ and certain
 other scales have been 
 integrated out. Then $M$ should be $M^{ }_\rmi{th}$ as defined above
 \eq\nr{neq_final}. In order to simplify the notation and because the
 thermal correction $\delta M^{ }_\rmi{th} = \Delta M^2_\rmi{th}/(2 M)$
 is numerically very small, we however keep the notation
 $M$ for the heavy-particle mass 
 in the following. In contrast, the Salpeter correction 
 of \eq\nr{Salpeter} is important; in our 
 approach it emerges ``dynamically''
 from the potential in \eq\nr{master}.   
 }  

Following \eqs(4.1)--(4.15) of ref.~\cite{peskin}, 
the spectral function can be extracted from the imaginary part of 
a ``Coulomb Green's function''. This Green's function satisfies 
an inhomogeneous  Schr\"odinger-type equation, 
with the feature that 
the static potential contains a Debye-screened real part, 
as well as an imaginary part
($\Gamma$ in the notation of \se\ref{se:physics}). 
The latter represents frequent
thermal scatterings on light plasma constituents that decohere the 
DM particles. The processes are illustrated
in \fig\ref{fig:processes}.

%%%%%%%%%%%%%%%%%%%%%%%%%% FIGURE %%%%%%%%%%%%%%%%%%%%%%%%%%%%%%%%%%%%%%%%%
%
\begin{figure}[t]
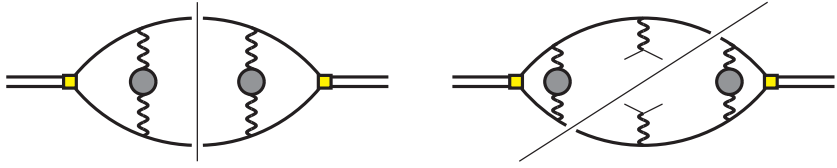


\hspace*{6mm}%
\begin{minipage}[c]{14.2cm}
\begin{eqnarray*}
&& 
 \hspace*{-1cm}
 \NRa 
 \hspace*{1.5cm}
 \NRb
\end{eqnarray*}
\end{minipage}

\vspace*{4mm}

\caption[a]{\small 
 Processes incorporated through the solution
 of the Schr\"odinger equation, \eq\nr{Seq}, with 
 a temperature-modified static potential, \eq\nr{master}.  
 The thin line indicates a cut 
 (i.e.\ an imaginary part, cf.\ \eq\nr{get_rho}). 
 The complete solution
 includes an infinite re-iteration of both types of processes. 
 Left: virtual corrections, originating from $V(r)$. 
 Right: real corrections, originating from $\Gamma(r)$.
 }
\la{fig:processes}
\end{figure}
%
%%%%%%%%%%%%%%%%%%%%%%%%%%%%%%%%%%%%%%%%%%%%%%%%%%%%%%%%%%%%%%%%%%%%%%%%%%

Let us define
\be
 E' \; \equiv \; \omega -2 M - \frac{k^2}{4M}
 \;, \la{Eprime}
\ee
where $k$ is the momentum of the \dmdm pair with respect to the heat
bath (cf.\ \eq\nr{rho_k}). Through a slight abuse of notation we now
redefine $\rho$ to stand for the spectral function related to relative
dynamics, $\rho(\omega,\vec{k}) \equiv \rho(E')$. 
A non-relativistic Hamiltonian is written as 
\be
 H = -\frac{\nabla_r^2}{M} + V(r) 
 \;, \la{H}
\ee
where $V(r)$ contains virtual corrections such as 
Debye screening and temperature-modified mixing angles. 
Then the spectral function is obtained from 
\ba
 \bigl[ H - i \, \Gamma(r) - E' \bigr] G(E';\vec{r},\vec{r'}) & = & 
 N\, \delta^{(3)}(\vec{r-r'}) \;, \la{Seq} \\ 
 \lim_{\vec{r,r'}\to \vec{0}} \im  G(E';\vec{r},\vec{r'})
 & = & \rho(E') \la{get_rho}
 \;, 
\ea
where $N$ is the number of degrees of freedom. 
Eq.~\nr{Seq} represents a Fourier transform of a time-dependent
Schr\"odinger equation with a local source created at $t=0$ 
and annihilated at time $t > 0$. The simplest realistic scenarios
contain a complex scalar field or a non-relativistic fermionic spinor, 
for which $N=2$. We note that in vacuum, i.e.\ by setting 
$\Gamma(r)\to 0^+$, the spectral function possesses the usual
quantum-mechanical interpretation, 
\be
 \lim_{T\to 0} \rho(E') = 
 N \sum_m |\psi^{ }_m(\vec{0})|^2 \pi\, \delta(E^{ }_m - E')
 \;, \la{qm}
\ee 
where $E^{ }_m$ are the $s$-wave 
energy eigenvalues related to \eq\nr{H} and 
$\psi^{ }_m$ are the corresponding wave functions. 

When expressed in the center-of-mass coordinates
of \eq\nr{Eprime}, the integral over $k$ can be carried out
in the Laplace transform of \eq\nr{Laplace}. We get
\ba
 \gamma & \approx &
 \int_{\vec{k}} e^{-\frac{2M}{T} - \frac{k^2}{4 M T}}
 \int_{-\Lambda}^{\infty} \! \frac{{\rm d} E'}{\pi} \, e^{-E'/T} \, \rho(E') 
 \nn
 & = & 
 \Bigl( \frac{MT}{\pi} \Bigr)^{3/2} e^{-2M/T}
 \int_{- \Lambda}^{\infty} 
 \! \frac{{\rm d}E'}{\pi} \, e^{-E'/T} \, \rho(E') 
 \;. \la{Laplace2}
\ea 
In the 
free limit, corresponding to $V(r)\to 0$ and $\Gamma\to 0^+$, 
\eqs\nr{Seq} and \nr{get_rho} yield 
\be
 \rho^{ }_{ }(E') \; \to \; 
 \rho^{ }_\rmi{free}(E') \; \equiv \; 
 \frac{N M^{\fr32}  \theta(E') \sqrt{E'} }{4\pi}   
 \;. \la{rho_free}
\ee
Combining \eqs\nr{Laplace2} and \nr{rho_free} and comparing with 
\eq\nr{neq_final}, we get
\be
 \gamma^{ }_\rmi{free} \approx 
 N \biggl[ 
 \Bigl( \frac{MT}{2\pi} \Bigr)^{3/2} 
 e^{-M/T}
 \biggr]^2
 \approx \frac{ n_\rmi{eq}^2 }{4 N} 
 \;. \la{Gamma_free}
\ee

A nice method to solve \eqs\nr{Seq} and \nr{get_rho} is to reduce the 
solution of the inhomogeneous equation into the solution of the corresponding 
homogeneous equation which is regular at origin~\cite{original}. 
Let $\rho \equiv \alpha M r$, 
$V \equiv \alpha^2 M \widetilde{V}$, 
$\Gamma \equiv \alpha^2 M\, \widetilde{\Gamma}$, 
$E' \equiv \alpha^2 M \widetilde{E}'$, and denote by $\ell$
an angular quantum number. 
Then the radial homogeneous equation takes the form
\be
 \biggl[
  - \frac{{\rm d}^2}{{\rm d}\rho^2} + \frac{\ell(\ell+1)}{\rho^2}
 + \widetilde{V} - i \widetilde{\Gamma} - \widetilde{E}'  
 \biggr] \, u^{ }_\ell(\rho) = 0 
 \;. \la{Schr_rescaled}
\ee
The regular solution is the one with the asymptotics 
$u^{ }_\ell = \rho^{\ell + 1}$ at $\rho \ll 1$. With this normalization, 
the $s$-wave spectral function is obtained from
\be
 \rho(E') \; = \; \frac{\alpha N M^2}{4\pi}
 \int_0^\infty \! {\rm d}\rho \, \im \biggl[
  \frac{1}{(u^{ }_0)^2} \biggr]
 \;. \la{rho_rescaled}
\ee

%%%%%%%%%%%%%%%%%%%%%%%%%%% SUBSECTION %%%%%%%%%%%%%%%%%%%%%%%%%%%%%%%%%%%
%
\subsection{Resummed gauge field propagator and static potential}
\la{ss:prop}

An essential role in the solution of \eq\nr{Schr_rescaled} is played
by the static potential $V$ and by its imaginary part, denoted by $-i \Gamma$.
In typical DM models,  
$
 \lim_{T\to 0} V = - \alpha e^{- m r}/r
$
and
$
 \lim_{T\to 0} \Gamma = 0
$. 
Then bound states exist if $M \gsim 1.6 m / \alpha$ 
(cf.,\ e.g.,\ ref.~\cite{wimpo4}).
In the regime where $\pi T \ll m$, thermal corrections are
exponentially suppressed and bound states are not affected. 
Once $\pi T \sim m$, thermal corrections are of order unity, 
however they are rather complicated in this regime; 
we do {\em not} consider this situation. Rather, we go over to 
temperatures $\pi T \gg m$, which for $Z$ boson exchange 
corresponds to $T \gg 30$~GeV.  Even if $\pi T$ is large
compared with $m$, it is still small compared with $M$, 
which is assumed to satisfy $M \gsim 20 T$.

In the regime $\pi T \gg m$, the gauge field self-energy obtains
the so-called Hard Thermal Loop (HTL) form~\cite{ht1,ht2,ht3,ht4}
(for a derivation, see appendix~A).\footnote{%
 There are also HTL vertex corrections, but for heavy  
 particles these can be omitted, cf.\ e.g.\ ref.~\cite{chesler}. 
 } 
This means that the gauge boson 
mass $m$ is modified by a thermal correction of 
order $gT$, which is parametrically of the same order as $m$, or larger. 
The self-energy is in general momentum-dependent, however 
the relevant momentum scale is $k \sim m \ll \pi T$. Therefore 
momentum dependence is 
suppressed by $\sim k^2/(\pi T)^2 \ll 1$. 

In a thermal system, several different self-energies can be defined, 
depending on the time ordering chosen. Only one choice  
can be consistently used in connection with \eq\nr{Schr_rescaled}.
Given that the \dmdm pair is heavy and therefore behaves
essentially as in vacuum, its
interactions with gauge fields are encoded in a
{\em time-ordered correlator}.
For completeness 
we show this explicitly around \eq\nr{proof}. 
At finite temperature
this result has previously been established (directly or indirectly)
in the context of QCD~\cite{static,jacopo} 
and QED~\cite{bbr}. 

The time-ordered propagator can be straightforwardly
determined within the so-called imaginary-time formalism, 
in which the Feynman rules are identical to those 
in vacuum, apart from a Wick rotation. Then we compute
an imaginary-time correlator, denoted by $\prop_{00\rmii{E}}$, 
for the temporal gauge field components with a Matsubara frequency $k_n$,  
and analytically continue it
to obtain a retarded correlator,
\be
 \prop_{00\rmii{R}} = \left. \prop_{00\rmii{E}} 
 \right|^{ }_{k_n \to -i [\omega + i 0^+]}
 \;. 
\ee
Subsequently the time-ordered propagator reads
(cf.\ e.g.\ refs.~\cite{mlb,book})
\be
 i \prop_{00\rmii{T}}(\omega,k)
 = \prop_{00\rmii{R}}(\omega,k)
 + 2 i \nB{}(\omega) \im \prop_{00\rmii{R}}(\omega,k) 
 \;. \la{Delta_T}
\ee
Given that for the static potential we are only interested in 
the static limit and that $\nB{}(\omega)\approx T/\omega$ for $\omega\ll T$, 
it is sufficient in practice to consider 
\be
 i \prop_{00\rmii{T}}(0,k)
 = \prop_{00\rmii{R}}(0,k)
 + i \lim_{\omega\to 0} \frac{2T}{\omega}
 \im \prop_{00\rmii{R}}(\omega,k)  
 \;. \la{Delta_T_limit}
\ee
The static potential and the thermal width are obtained from 
(cf.\ \se\ref{se:nondeg})
\be
 V(r) - i\, \Gamma(r) = g^2\CF  
 \int \! \frac{{\rm d}^3\vec{k}}{(2\pi)^3}
 \, 
 \Bigl( 1 - e^{i\vec{k}\cdot\vec{r}} \Bigr)
 \, i \prop_{00\rmii{T}}(0,{k}) 
 \; - \; \delta V
 \;, \la{master} 
\ee
where we assume the counterterm $\delta V$ to
be so chosen that $\lim_{r\to\infty} V(r) = 0 $ at $T = 0$.\footnote{%
 More precisely, a counterterm is needed because the 
 self-energy correction is linearly ultraviolet divergent in vacuum. 
 Its finite part defines what we mean by the 
 renormalized heavy-particle mass $M$. 
 } 
The $\vec{r}$-independent part originates from self-energy corrections
and the $\vec{r}$-dependent one from exchange contributions, 
and $\CF$ is a Casimir factor. 

The real and imaginary parts of the gauge field propagator, 
\eq\nr{Delta_T_limit}, lead to specific physical phenomena which have 
been illustrated in \fig\ref{fig:processes}. The real part 
corresponds to ``virtual exchange'', i.e.\ a Debye screened potential. 
The imaginary part corresponds to ``real scatterings'', specifically
the scattering of the heavy particles on light plasma constituents; 
its physical origin is reiterated
in \eqs\nr{imag1} and \nr{imag2}. 
For $r\to \infty$, 
$V(\infty)$ corresponds to twice the heavy particle thermal mass 
correction (cf.\ \eq\nr{Salpeter}), 
and $\Gamma(\infty)$ to twice the heavy particle thermal 
interaction rate~\cite{ht1,bbr}. 
The interpretation of $\Gamma$ in the language of open quantum
systems has been discussed in ref.~\cite{ya}. 
Finally, we recall that the Bose-enhanced term in \eq\nr{Delta_T_limit}, 
representing large occupation numbers $\sim T/\omega \gg 1$, 
has a classical plasma physics interpretation: electric fields 
exert a Lorentz force on charged particles, which 
induces a current, by which the electric field is reduced. 
In the real-time formalism, the Bose-enhanced contribution originates
from the 
$rr$-propagator in the $r/a$ basis, 
and gives the dominant contribution to typical
soft observables~\cite{sch}. 

%%%%%%%%%%%%%%%%%%%%%%%%%%% SUBSECTION %%%%%%%%%%%%%%%%%%%%%%%%%%%%%%%%%%%
%
\subsection{Summary of the theoretical framework}

We have argued that the computation of massive dark matter relic 
density can be factorized into a number of independent steps. 
First, the thermal self-energies of the particles exchanged by the 
dark ones need to be computed. From the 
self-energies, the corresponding propagators can 
be determined (cf.\ \eq\nr{Delta_T_limit}). These fix the 
static potential and the thermal width experienced by the annihilating
pair (cf.\ \eq\nr{master}). Subsequently the spectral function can 
be computed through the solution of a Schr\"odinger equation
(cf.\ \eqs\nr{Seq} and \nr{get_rho}). Its Laplace-transform
gives the thermally averaged annihilation rate (cf.\ \eq\nr{Laplace2}). 
The annihilation rate parametrizes a rate equation, which can be 
integrated to give the final non-equilibrium number density 
(cf.\ \eq\nr{linear_response} or \nr{Boltzmann}). In principle the 
uncertainties of each of these steps can be scrutinized and 
improved upon separately. 

%%%%%%%%%%%%%%%%%%%%%%%%%%% SECTION %%%%%%%%%%%%%%%%%%%%%%%%%%%%%%%%%%%%%%
%
\section{$Z$ exchange at finite temperature}
\la{se:Z}

Our first physics goal is to 
apply the formalism of \se\ref{se:theory} to determine
the spectrum of a kinetically equilibrated
\dmdm pair interacting through $Z$ boson exchange.
Non-relativistic particles 
interacting with $Z$ bosons are represented either by a complex scalar field
or by a two-component spinor, and in general the two degrees of freedom have 
different masses. Here we focus on a case in which
the two degrees of freedom are degenerate in mass; the non-degenerate
case is addressed in \se\ref{se:nondeg}. 

With this setup, the parameters defined in \se\ref{ss:prop} are 
\be
 \alpha \; \equiv \; \frac{ g_1^2 + g_2^2}{ 16\pi} 
        \; \approx \; 0.01
 \;, \quad
 m \; \equiv \; \mZ \; \approx \; 91~\mbox{GeV}
 \;, 
\ee 
where $g^{ }_1$ and $g^{ }_2$ are the hypercharge and weak gauge 
couplings, respectively. Solving a static Schr\"odinger equation
with a Yukawa potential with these parameters, 
a $1s$ bound state is found for $M \gsim 15$~TeV. 
Here we consider $M \lsim 10$~TeV so that no bound states exist.\footnote{%
 For completeness we note that if $M \gsim 15$~TeV, bound states
 exist but they melt 
 at temperatures below the thermal freeze-out, in analogy with the 
 case of $Z'$ exchange considered in \fig\ref{fig:Z0p}. 
 } 

The general forms of 
the self-energies and propagators needed for $Z$ exchange are 
reviewed in appendix~A. Here we proceed with the propagator
from \eq\nr{Z0_prop}. We denote the vacuum and thermal mixing angles 
by $\theta$ and $\tilde\theta$, where~\cite{broken}
\ba
 \sin(2\theta^{ }_\rmii{}) 
 & \equiv & \frac{2g^{ }_1 g^{ }_2}{g_1^2 + g_2^2}
 \;, \\ 
 \sin(2\tilde\theta) 
 & \equiv &
 \frac{\sin(2\theta) \mZ^2 }{\sqrt{\sin^2(2\theta) \mZ^4 +
 [\cos(2\theta) \mZ^2 + m_\rmii{E2}^2 - m_\rmii{E1}^2]^2}}
 \;. \la{mixing} 
\ea
The U$^{ }_\rmii{Y}$(1) and SU$^{ }_\rmii{L}$(2) Debye masses read~\cite{meg}
\be
 m^{2}_\rmii{E1} \; \equiv \; 
 \Bigl( \fr{\nS}6 + \frac{5\nG}{9} \Bigr) g_1^2 T^2 
 \;, \quad
 m^{2}_\rmii{E2} \; \equiv \; 
 \Bigl( \fr23 + \fr{\nS}6 + \frac{\nG}{3} \Bigr) g_2^2 T^2
 \;, \la{Debye}
\ee
where $\nS \equiv 1$ and $\nG \equiv 3$ are the numbers 
of Higgs doublets and fermion generations, respectively.
The neutral eigenstates have the masses
\ba
  \mZt^2 & \equiv &  m_{+}^2 
  \;, \quad
  \mQt^2 \; \equiv \;  m_{-}^2  
  \;, \la{mZt_mQt} \\ 
  m_{\pm}^2 & \equiv & 
 \frac{1}{2} 
 \Bigl\{
   \mZ^2 + m_\rmii{E1}^2 + m_\rmii{E2}^2 \pm 
   \sqrt{\sin^2(2\theta) \mZ^4 +
   [\cos(2\theta) \mZ^2 + m_\rmii{E2}^2 - m_\rmii{E1}^2]^2}
 \Bigr\}
 \;. 
\ea
Then the potential from \eq\nr{master}, fixing $\delta V$
from $\lim_{r\to \infty}\lim_{T\to 0}V(r) = 0$, takes the form
\be
 V(r) \approx 
 \alpha \, \biggl\{ 
   \mZ^{\rmii{$(T=0)$}} - \cos^2(\tilde\theta - \theta) 
%   \biggl(\mZt + \frac{e^{-\mZt r}}{r} \biggr)
   \biggl[\mZt + \frac{\exp({-\mZt r})}{r} \biggr]
   - \sin^2(\tilde\theta - \theta) 
%   \biggl(\mQt + \frac{e^{-\mQt r}}{r} \biggr)
   \biggl[\mQt + \frac{\exp({-\mQt r})}{r} \biggr]
 \biggr\}
 \;, \la{Vr}
\ee 
whereas the imaginary part can be expressed as
\ba
 \Gamma(r) & \approx & 
 \alpha T \, \biggl\{ 
 \frac{ \cos^2(\tilde\theta - \theta) 
 (m_\rmii{E1}^2 \sin^2\tilde{\theta}  + m_\rmii{E2}^2 \cos^2\tilde{\theta} )
 \, \phi(\mZt r)}{\mZt^2}
 \nn & & \quad + \, 
  \frac{ \sin^2(\tilde\theta - \theta) 
 (m_\rmii{E1}^2 \cos^2\tilde{\theta}  + m_\rmii{E2}^2 \sin^2\tilde{\theta} )
 \, \phi(\mQt r)}{\mQt^2}
 \nn & & \quad + \, 
 \frac{\sin(2(\tilde{\theta}-\theta))\sin(2\tilde{\theta})
 (m_\rmii{E2}^2 - m_\rmii{E1}^2)\, \theta(\mZt r,\mQt r) }
 {2(\mZt^2 - \mQt^2)}
 \biggr\} 
 \;. \la{Gammar}
\ea
Here we have defined
\ba
 \phi(\rho) & \equiv & 
 1 - 
 2 \int_0^\infty \! \frac{{\rm d}x }{(x^2+1)^2}
 \, \frac{\sin(x \rho)}{\rho} 
 \;, \la{phi} \\  
 \theta(\rho^{ }_1,\rho^{ }_2) & \equiv & 
 2 \ln \biggl( \frac{\rho^{ }_1}{\rho^{ }_2} \biggr) + 
 2 \int_0^\infty \! \frac{{\rm d}x }{x^2+1}
 \, \biggl[ \frac{\sin(x \rho^{ }_1)}{\rho^{ }_1}
 - \frac{\sin(x \rho^{ }_2)}{\rho^{ }_2}
 \biggr]
 \;, \la{theta}
\ea
both of which vanish at zero separation ($r\to 0$).

We consider a semi-realistic choice for the dark matter
mass scale, $M\gsim 1$~TeV. As alluded to above, the lower
bound is dictated by the 
ease of computation, but 
phenomenological constraints from the LHC 
favour a similar value. 
% Coupling constants and  masses 
% are fixed as explained in appendix~E of ref.~\cite{broken}. 
Results are shown in \fig\ref{fig:Z0}. 
The thermal scatterings experienced by the DM particles 
with light plasma constituents (cf.\ \fig\ref{fig:processes})
cause the 2-particle threshold to smoothen, but
the most important effect is related to 
Debye screening, both through the shift of the threshold
location according to the Salpeter correction from \eq\nr{Salpeter}
and through modified Sommerfeld factors, as we now explain.  

%%%%%%%%%%%%%%%%%%%%%%%%%%%%%%%%% FIGURE %%%%%%%%%%%%%%%%%%%%%%%%%%%%%%%%%
\begin{figure}[t]

\hspace*{-0.1cm}
\centerline{%
% \epsfxsize=7.5cm\epsfbox{Z_M20.eps}
% \hspace{0.1cm}
 \epsfxsize=8.8cm\epsfbox{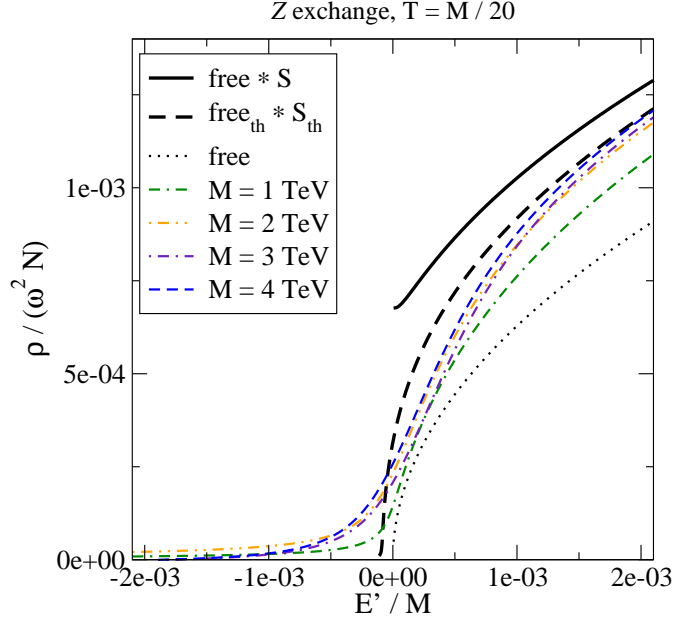}%
}

\caption[a]{\small
 The free (dotted line; cf.\ \eq\nr{rho_free}) and resummed
 (coloured lines; cf.\ \eq\nr{rho_rescaled})  spectral 
 functions, with $E'$
 denoting the energy with respect to the 2-particle threshold
 and $\omega \equiv 2 M + E'$. 
 The potential and width are 
 from \eqs\nr{Vr} and \nr{Gammar}, respectively. 
 The free result multiplied
 by the Sommerfeld factor~$S$ from \eq\nr{Ss} is 
 shown with a solid line. We also 
 display a numerically determined Sommerfeld factor which 
 includes the effects of  
 Debye screening and a shift of the threshold location 
 according to the Salpeter correction from \eq\nr{Salpeter}
 (free$_\rmi{th} {\raise-0.5ex \hbox{*}} S_\rmi{th}$; $M = 3$~TeV). 
}

\la{fig:Z0}
\end{figure}
%%%%%%%%%%%%%%%%%%%%%%%%%%%%%%%%%%%%%%%%%%%%%%%%%%%%%%%%%%%%%%%%%%%%%%%%%%%

In \fig\ref{fig:Z0} we show with a solid line the result corresponding to 
a Sommerfeld factor for attractive Coulomb exchange. 
This can be expressed as~\cite{fadin}
\be
  \Ss = \frac{ \Xs } { 1 - e ^{ - \Xs } }
  \;, \quad 
  \Xs   =  \frac{ \pi\alpha  } { v }
  \;, \label{Ss}
\ee
where $E'$ from \eq\nr{Eprime} has been parametrized through a velocity as 
$E' = M v^2$. We use the Coulomb form, because for $M \gsim 3$~TeV electroweak
symmetry is restored around the freeze-out temperature. 
The main difference from the Coulomb case 
is due to Debye screening (cf.\ \eq\nr{Debye}), which persists
at high temperatures. The numerically determined 
Debye-screened Sommerfeld factor  
has been illustrated in \fig\ref{fig:Z0} with a dashed line, 
and agrees well with the full solution soon above the threshold. 

The total annihilation rate $\gamma$ is given by the Laplace transform
in \eq\nr{Laplace2}. Given that $M/T \sim 20 ... 25$, the Laplace transform 
corresponds to an average over the range $E'/M \lsim  0.1$. 
This is a broad range in comparison with the threshold region
$|E'| \lsim 20 \alpha^2 M$ shown in \fig\ref{fig:Z0}. 
In particular, the suppression
with respect to the Debye-screened 
Sommerfeld prediction at $E' > 0$ is largely
compensated for by the enhanced spectral weight at $E'<0$. 
Moreover, the suppression of the Sommerfeld factor by
Debye screening amounts to a higher-order correction to the total rate. 
The shift
of the threshold location to the left increases the annihilation rate
according to \eq\nr{Salpeter}, but the suppression of the 
Sommerfeld factor by Debye screening decreases it; we find 
that the final result for $\gamma$ is $\sim 1$\% below
the Coulombic Sommerfeld estimate.\footnote{%
 Thermal corrections were anticipated to be small in ref.~\cite{mbx}, 
 however only some of them were included. 
 } 
The enhancement with 
respect to the free result is $\sim 9$\%.

%%%%%%%%%%%%%%%%%%%%%%%%%%% SECTION %%%%%%%%%%%%%%%%%%%%%%%%%%%%%%%%%%%%%%
%
\section{$Z'$ exchange at finite temperature}
\la{se:Zp}

For a further illustration we
move on to a technically simpler model, similar to those for 
which ``wimponium'' bound states were found at 
zero temperature~\cite{wimpo1,wimpo2,wimpo3,wimpo4}. More specifically, 
we consider a dark sector with an U(1) gauge symmetry, coupled  
to the Standard Model through a vector or Higgs portal
(cf.\ e.g.\ refs.~\cite{gamma1,gamma2,gamma3}).  
Being only interested in qualitative features the portal couplings
will be omitted for practical purposes, apart from assuming that
the dark sector is in kinetic equilibrium with the Standard Model. 
The dark sector then consists
of the heavy dark matter particle ($\psi$), the dark gauge boson 
($V^{ }_\mu$), and a dark Higgs field ($S$) which 
gives the dark gauge boson a mass $\mV \gsim 1$~GeV as is required for 
phenomenology (cf.\ e.g.\ refs.~\cite{gamma5,gamma6}). 
We refer to the dark gauge boson as $Z'$. 

A concrete realization of the above setup is provided 
by the Lagrangian 
\be
 \mathcal{L} = \mathcal{L}^{ }_\rmii{SM} 
 + \mathcal{L}_\rmi{portal}
 -\fr14 V^{\mu\nu}_{ }  V^{ }_{\mu\nu}
 + (D_{ }^\mu S)^* (D^{ }_\mu S) - V(S^*\! S)
 + \bar{\psi}(i \gamma^\mu D^{ }_\mu - M )\psi
 \;, \quad
\ee
where 
$V^{ }_{\mu\nu}$ is the field strength corresponding to the dark U(1).
The potential breaks the U(1) gauge symmetry spontaneously, 
$V(S^*\!S) \simeq -\nu^2 S^*\!S + \lambda' (S^*\!S)^2$, 
$\nu^2,\lambda' > 0$.
Portal couplings have the form 
$
 \mathcal{L}_\rmi{portal} =
  - \kappa^{ }_1  V_{ }^{\mu\nu} F^{ }_{\mu\nu}
  - \kappa^{ }_2 S^*\!S H^\dagger H
$,  
where $F^{ }_{\mu\nu}$ is the Standard Model hypercharge
field strength and $H$ is the Higgs doublet.
Both $\kappa^{ }_1$ and $\kappa^{ }_2$ 
are assumed small enough to be insignificant in practice. 
The coupling associated with the dark U(1) group
is denoted by $e'$, and $D^{ }_\mu = \partial^{ }_\mu - i e' V^{ }_\mu$. 
In accordance with ref.~\cite{wimpo1}, in which
the phenomenology of this model was discussed, we take 
$\alpha' \equiv (e')^2 / (4\pi) \sim 0.01$. The mass of the scalar
particle is assumed to be $\mS \sim 1$~GeV but it plays little role. 
Dark matter particles with $M \sim$~TeV freeze 
out in the non-relativistic regime as usual.
For these parameters the constraint $M \gsim 1.6 \mV/\alpha'$ 
(cf.,\ e.g.,\ ref.~\cite{wimpo4}) is well 
satisfied, guaranteeing the existence of bound states in vacuum.

%%%%%%%%%%%%%%%%%%%%%%%%%%%%%%%%% FIGURE %%%%%%%%%%%%%%%%%%%%%%%%%%%%%%%%%
\begin{figure}[t]

\hspace*{-0.1cm}
\centerline{%
 \epsfxsize=8.8cm\epsfbox{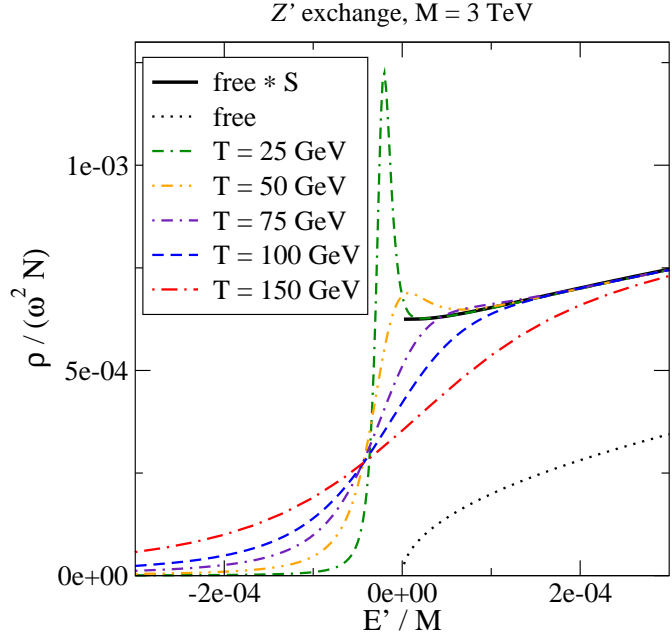}%
% \hspace{0.1cm}
% \epsfxsize=7.5cm\epsfbox{Z_Zp_scan.eps}
}

\caption[a]{\small
 The free, Sommerfeld enhanced, and resummed spectral functions
 for $Z'$ exchange,  
 for $M = 3$~TeV and $N=2$ and the potential and width from 
 \eq\nr{Gammarp}.
 The notation is as in \fig\ref{fig:Z0}.
}

\la{fig:Z0p}
\end{figure}
%%%%%%%%%%%%%%%%%%%%%%%%%%%%%%%%%%%%%%%%%%%%%%%%%%%%%%%%%%%%%%%%%%%%%%%%%%%

A computation of the $Z'$ self-energy in this model 
is presented in appendix~B. 
Defining a thermally modified $Z'$ mass as
\be
 \mVt^2 \; \equiv \;  \mV^2 + m_\rmii{E$'$}^2  
 \;, \quad 
 \mV^2 \; \equiv \; e'{}^2\vT'{}^2
 \;, \quad
 m_\rmii{E$'$}^2 \; \equiv \; \frac{e'{}^2 T^2}{3}
 \;, \la{mDebyep}
\ee
where $\vT'$ is the thermal expectation value of $S$
($S = {\vT'}/{\sqrt{2}} + ...$), 
and choosing $\delta V$ so that 
$\lim_{r\to \infty}V(r) = 0$ at $T=0$, 
\eqs\nr{DeltaTp} and \nr{master} yield
\ba
 V(r) & \approx &  
 \alpha'  \biggl\{ 
 \mV^{\rmii{$(T=0)$}} - \biggl[ \mVt
 + \frac{\exp({-\mVt r})}{r} \biggr] \biggr\}
 \;, \quad % \la{Vrp} \\ 
 \Gamma(r) \; \approx \; 
 \frac{\alpha' T\, m_\rmii{E$'$}^2 \, \phi(\mVt r)}{\mVt^2} 
 \;. \hspace*{5mm} \la{Gammarp}
\ea
Here the function $\phi$ is from \eq\nr{phi}. 
The spectral function is determined from \eqs\nr{Schr_rescaled}
and \nr{rho_rescaled}. Given that $\pi T \gg \mV$, the 
mass $\mV$ is insignificant in practice; in fact the dark U(1) symmetry 
is restored at the temperatures at which freeze-out takes place. 

Illustrative results for $M = 3$~TeV 
are shown in \fig\ref{fig:Z0p}. The bound state peak is found to 
dissolve at a temperature $T\sim 75$~GeV, i.e.\ below 
freeze-out, $T^{ }_\rmi{freeze-out} \gsim 100$~GeV.
The spectral function gets smoothened across the threshold.
Around $T^{ }_\rmi{freeze-out}$  
the physical annihilation rate obtained from the Laplace transform
in \eq\nr{Laplace2} is however in good agreement 
with that predicted by the Sommerfeld factor.

%%%%%%%%%%%%%%%%%%%%%%%%%%% SECTION %%%%%%%%%%%%%%%%%%%%%%%%%%%%%%%%%%%%%%
%
\section{Gluon exchange at finite temperature}
\la{se:gluon}

Let us turn to strong interactions. In the context of supersymmetric 
theories, one scenario that has attracted interest is the case of neutralino
dark matter, which could co-annihilate with gluinos just slightly heavier
than neutralinos. The gluinos themselves may form
bound states, which also annihilate. This system has
been analyzed within a Boltzmann equation approach in, 
for instance, refs.~\cite{old5,old51}. (Much the same could be done if
gluinos were replaced by stops, 
cf.~e.g.~refs.~\cite{stop1,stop15,stop2} and references therein.)

For the purposes of the present paper, we only consider the gluino part of 
the set of non-equilibrium variables.\footnote{%
  The full set of dark matter rate equations necessitates a non-trivial
  discussion in the co-annihilation regime, implying in particular that
  the gluino annihilation rate contributes to an effective
  $\langle \sigma v\rangle$ with a certain weight but is not
  the only ingredient~\cite{old1}. For simplicity we discuss here only the 
  gluino annihilation rate. 
 } 
The question is whether gluino bound 
states can persist up to high temperatures and, if so, how strongly they would
affect the gluino annihilation rate. 

%%%%%%%%%%%%%%%%%%%%%%%%%%%%%%%%% FIGURE %%%%%%%%%%%%%%%%%%%%%%%%%%%%%%%%%
\begin{figure}[t]

\hspace*{-0.1cm}
\centerline{%
 \epsfxsize=8.8cm\epsfbox{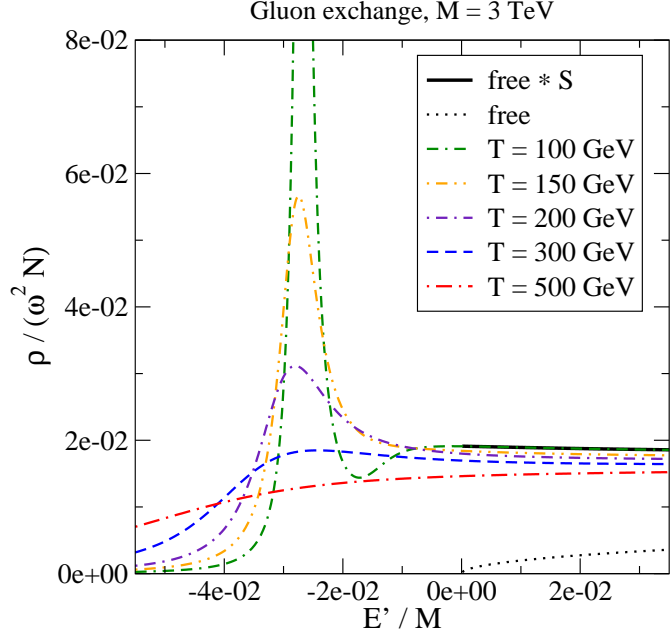}%
% \hspace{0.1cm}
% \epsfxsize=7.5cm\epsfbox{Z_Zp_scan.eps}
}

\caption[a]{\small
 The free and resummed  spectral functions for gluon exchange, 
 with $M = 3$~TeV and $N=16$. The potential and width are from 
 \eq\nr{Gammarg}. 
 The notation is as in \fig\ref{fig:Z0}.
 The Sommerfeld factor $S$ was computed for $T = 100$~GeV.
}

\la{fig:gluon}
\end{figure}
%%%%%%%%%%%%%%%%%%%%%%%%%%%%%%%%%%%%%%%%%%%%%%%%%%%%%%%%%%%%%%%%%%%%%%%%%%%

For the case of gluon exchange, the results for the real-time 
static potential can be taken over from  
QCD literature~\cite{static,bbr,jacopo}, with a simple change
of group theory factors.
Defining the Debye mass and an effective coupling for the 
adjoint matter representation as 
\be
 m_\rmii{E3}^2 \; \equiv \; 
 \Bigl( 1 + \frac{\nG}{3} \Bigr) g_3^2 T^2
 \;, \quad 
 \alpha^{ }_3 \; \equiv \;  \frac{3g_3^2}{4\pi} 
 \;, \la{mDebyeg}
\ee
where $\nG = 3$ is the number of generations and $g^2_3 \equiv 4\pi \alphas$, 
and concentrating on the attractive interaction in the singlet channel 
like in ref.~\cite{old5}, we get
\ba
 V(r) & \approx &  
 - \alpha^{ }_3 
 \biggl[ m_\rmii{E3}^{ }
 + \frac{\exp({- m_\rmii{E3}^{ } r})}{r} \biggr] 
 \;, \quad % \la{Vrg} \\ 
 \Gamma(r) \; \approx \; 
 \alpha^{ }_3 T\, \phi ( m_\rmii{E3}^{ } r ) 
 \;, \la{Gammarg}
\ea
where $\phi$ is from \eq\nr{phi}. Here $V(\infty)$ and $\Gamma(\infty)$
correspond to a thermal mass correction (cf.\ \eq\nr{Salpeter}) 
and interaction rate~\cite{ht1} of two independent heavy gluinos.

%%%%%%%%%%%%%%%%%%%%%%%%%%%%%%%%% FIGURE %%%%%%%%%%%%%%%%%%%%%%%%%%%%%%%%%
\begin{figure}[t]

\hspace*{-0.1cm}
\centerline{%
% \epsfxsize=7.8cm\epsfbox{Zp_M3.eps}%
% \hspace{0.1cm}
 \epsfxsize=8.2cm\epsfbox{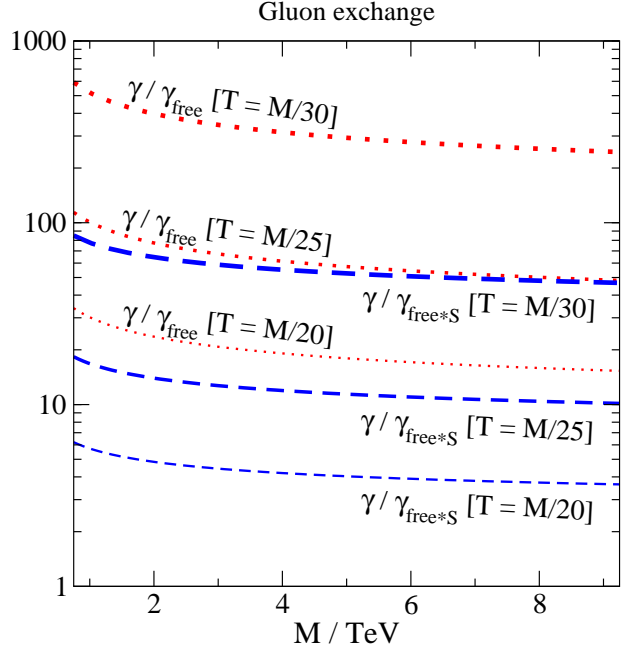}
}

\caption[a]{\small
 The ratio of the total rate from \eq\nr{Laplace2} to the 
 rates obtained from the approximations ``free'' and 
 ``free$\,*\, S$'', where $S$ denotes the Sommerfeld factor. 
 The spectral function originates from 
 gluon exchange and was illustrated in 
 \fig\ref{fig:gluon}. The total rate exceeds
 the Sommerfeld estimate
 by a factor $\sim 4...80$, and the free rate
 by $15 .... 600$, depending on parameter values. 
}

\la{fig:scan}
\end{figure}
%%%%%%%%%%%%%%%%%%%%%%%%%%%%%%%%%%%%%%%%%%%%%%%%%%%%%%%%%%%%%%%%%%%%%%%%%%%

The result of this procedure is shown in \fig\ref{fig:gluon}, 
for $M = 3$~TeV. 
% For a transparent numerical setup, we let the gauge coupling evolve
% at 1-loop level as 
% $g_3^2 = 8\pi^2 / [7 \ln(\bmu/\Lambdamsbar)]$, 
% where $\Lambdamsbar$ is fixed from the value 
% $\alphas(\mZ) = 0.1186$ and $\bmu = \pi T$.
At $T = M/20 = 150$~GeV, a bound state is clearly 
visible, and at $T = M/25$ even more so. 
If the gluino is substantially heavier than 
the DM particle then, for a given gluino mass $M$, the freeze-out  
temperature would be lower than $M/25$, and bound states would be 
very prominent.  

Once the temperature is high enough, bound states do dissolve even with strong
interactions. For instance, the curve $T = 500$~GeV in fig.~\ref{fig:gluon}
only shows a broad gradually rising spectral shape. Its general position is
shifted to the left of the free threshold because of the Salpeter correction
discussed below \eq\nr{Salpeter}. Because of frequent elastic scatterings with
plasma particles, which decohere any sharp quantum-mechanical features, the
spectral function is a smooth function. Bound states have disappeared because
of two reasons: Debye screening makes the potential less binding~\cite{ms}
and, already at a lower temperature, 
the thermal interaction rate (or width) caused by
the frequent elastic scatterings becomes larger than the binding energy of any
of the bound states~\cite{soto,review,wu}.

Integrating over the spectral function with the Boltzmann weight yields
the total annihilation rate, cf.\ \eq\nr{Laplace2}. The corresponding
results are shown in \fig\ref{fig:scan}, in comparison with  
results obtained from non-interacting ($\equiv \gamma^{ }_\rmi{free}$)
and Sommerfeld-enhanced ($\equiv \gamma^{ }_\rmi{free$\ast S$}$)
computations. Compared with the Sommerfeld-enhanced computation, the 
bound state contribution boosts the annihilation rate by a factor $4...80$, 
depending on parameter values.

%%%%%%%%%%%%%%%%%%%%%%%%%%% SECTION %%%%%%%%%%%%%%%%%%%%%%%%%%%%%%%%%%%%%%
%
\section{Non-degenerate masses}
\la{se:nondeg}

%%%%%%%%%%%%%%%%%%%%%%%%%% FIGURE %%%%%%%%%%%%%%%%%%%%%%%%%%%%%%%%%%%%%%%%%
%
\begin{figure}[t]
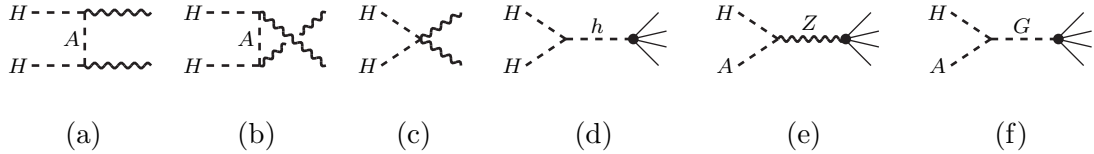


\hspace*{6mm}%
\begin{minipage}[c]{15.0cm}
\begin{eqnarray*}
&& 
 \hspace*{-1.2cm}
 \Deca
 \hspace*{0.0cm}
 \Decb
 \hspace*{0.0cm}
 \Decc
 \hspace*{-0.4cm}
 \Decd
 \hspace*{-0.2cm}
 \Dece
 \hspace*{-0.2cm}
 \Decf
\\[6mm]
&& \hspace*{-3mm}
   \mbox{(a)}
   \hspace*{1.8cm}
   \mbox{(b)}
   \hspace*{1.6cm}
   \mbox{(c)}
   \hspace*{1.9cm}
   \mbox{(d)}
   \hspace*{2.3cm}
   \mbox{(e)}
   \hspace*{2.3cm}
   \mbox{(f)}
\end{eqnarray*}
\end{minipage}

\vspace*{1mm}

\caption[a]{\small 
 Examples of how dark matter particles described
 by the model of \eq\nr{L_idm} can get annihilated. Wiggly lines stand for
 $Z$ bosons and thin solid lines for generic Standard Model particles. 
 The operators of \eq\nr{4ops}
 originate from processes (a)--(c) and their interference terms. 
 Process (d) mediated by the Higgs boson $h$ 
 and its interference with (a)--(c) is 
 numerically important (cf.\ e.g.\ ref.~\cite{schannel}) 
 but leads to no new operators. 
 All the reactions also take
 place with the exchange $H \leftrightarrow A$.
 Process (e) leads to a $p$-wave operator or to 
 effects suppressed by the mass difference $(\Delta M)^2$; 
 the latter type can also originate from 
 process (f) mediated by a Goldstone mode $G$. 
 }
\la{fig:decays}
\end{figure}
%
%%%%%%%%%%%%%%%%%%%%%%%%%%%%%%%%%%%%%%%%%%%%%%%%%%%%%%%%%%%%%%%%%%%%%%%%%%

We now proceed to cases, relevant e.g.\ for weak interactions, in which 
the particles interacting through gauge exchange are non-degenerate
in mass. We denote the mass difference by $\Delta M$.
If $\Delta M$ originates
from a Higgs mechanism, we expect it to be ``small'' in general, 
$\Delta M \lsim \mZ$. We work in a regime $\mZ \ll \pi T$
(cf.\ \se\ref{ss:prop}). Then $\Delta M \ll \pi T \ll M$, and 
the effects of $\Delta M$ can be incorporated within a non-relativistic
framework. Our goal is to show that having $\Delta M > 0$ changes the
situation only ``smoothly'' compared with the degenerate case. To this end
we consider a simple model and carry out a quantum-statistical
computation of correlators of the type illustrated in \fig\ref{fig:exchange}.

%%%%%%%%%%%%%%%%%%%%%%%%%%% SUBSECTION %%%%%%%%%%%%%%%%%%%%%%%%%%%%%%%%%%%
%
\subsection{A model and its non-relativistic description}

Consider a dark sector consisting of an additional
Higgs doublet. In the presence of electroweak symmetry breaking, there are
four physical states in this sector, the neutral ones denoted by $H$ and $A$
and two charged ones denoted by $H^\pm$. For simplicity we consider a situation
in which $m^{ }_\rmii{$H^\pm$} \gg \mH^{ }, \mA^{ }$. 
The state $H$ is taken to 
be the lightest particle ($M \equiv \mH^{ }$) and 
$A$ is slightly heavier ($\Delta M \equiv \mA^{ } - \mH^{ } > 0$).
The Lagrangian describing the interactions of these fields with 
physical $Z$ bosons reads
\be
 \mathcal{L} \; \equiv \; 
 \fr12 D^*_\mu (H + i A) D_{ }^\mu (H - i A) 
 - \fr12 \mH^2 % M^2 
         H^2
 - \fr12 \mA^2 % (M+ \Delta M)^2 
         A^2 + \ldots
 \;,  \la{L_idm}
\ee
where $D^{ }_\mu \equiv \partial^{ }_\mu + i g Z^{ }_\mu$ and 
$g \equiv \fr12 \sqrt{g_1^2 + g_2^2}$. There are also 
interactions with the Higgs boson (cf.\ \fig\ref{fig:decays}(d)) 
but these do not change the qualitative behaviour, so 
we omit them here. 

For a transparent discussion, it is helpful to 
go over into a non-relativistic Hamiltonian description. 
The interaction part of \eq\nr{L_idm} reads
\be
 \mathcal{L}^{ }_\rmi{int} =  
  g Z^\mu (A\, \partial^{ }_\mu H - H \partial^{ }_\mu A )
  + \frac{g^2}{2} Z^\mu Z^{ }_\mu (H^2 + A^2) + 
 \ldots
 \;. \la{Lint}
\ee
Key steps of the argument can be simplified by assuming
the scalar fields $H$ and $A$ to be so heavy
that they are essentially static; then they can be described 
by the non-relativistic modes $\phi$ and $\chi$ as
\be
 H \simeq \frac{1}{\sqrt{2\mH^{ }}}
 \Bigl( \phi\, e^{-i \mH^{ } t} + \phi^\dagger e^{i \mH^{ } t} \Bigr) 
 \;, \quad
 A \simeq \frac{1}{\sqrt{2\mA^{ }}}
 \Bigl( \chi\, e^{-i \mA^{ } t}
  + \chi^\dagger e^{i \mA^{ } t} \Bigr) 
 \;. \la{fields}
\ee
Inserting these decompositions into \eq\nr{Lint}; 
taking the limit $\mH,\mA\sim M \gg \mZ, \Delta M$; 
and defining subsequently an interaction Hamiltonian 
as $\mathcal{H}_\rmi{int} \equiv - \mathcal{L}_\rmi{int}$, we get  
\be
 \mathcal{H}_\rmi{int} = i g Z^{ }_0
 \Bigl( 
   \chi^\dagger \phi\, e^{i \Delta M t} - \phi^\dagger\chi\, e^{-i \Delta M t}
 \Bigr)
 + \rmO\Bigl( \frac{1}{M} \Bigr)
 \;. \la{Hint}
\ee
It is furthermore convenient to employ Euclidean (imaginary-time)
conventions for $Z^{ }_0$, i.e.\ $Z_0^M = i Z_0^E$; we use $Z_0^E$ in the
following, without displaying the superscript. Thereby the interaction
Hamiltonian between the static scalar fields and $Z$ bosons becomes
\be
 H_\rmi{int}(t) = - g \int_\vec{x} Z^{ }_0(t,\vec{x})
 \Bigl[ 
   (\chi^\dagger \phi)(\vec{x})\, e^{i \Delta M t} 
 - (\phi^\dagger\chi)(\vec{x})\, e^{-i \Delta M t}
 \Bigr]
 + \rmO\Bigl(\frac{1}{M}\Bigr)
 \;.  \la{Hint2}
\ee

Next, we need the four-particle operators describing the annihilations
of $H$ and $A$. Examples of processes are illustrated
in \fig\ref{fig:decays}. Considering only effects 
from the gauge vertices in \eq\nr{L_idm}, processes
(a)--(c) and their interference terms yield an imaginary
four-particle operator in the sense of ref.~\cite{bodwin}, 
\be
 \delta \mathcal{L}_\rmi{eff} 
 \; \simeq \; \frac{i g^4}{64\pi}
 \biggl( 
    \frac{\phi^\dagger \phi^\dagger \phi\, \phi}{\mH^2}
  \; + \; 
    \frac{\chi^\dagger \chi^\dagger \chi\, \chi}{\mA^2}
 \biggr)
 \;. \la{4ops}
\ee

%%%%%%%%%%%%%%%%%%%%%%%%%%% SUBSECTION %%%%%%%%%%%%%%%%%%%%%%%%%%%%%%%%%%%
%
\subsection{Derivation of a real-time static potential}
\la{ss:derivation}

Now, in accordance with the discussion in \se\ref{ss:rho}, 
the role of \eq\nr{4ops} is that it dictates the spectral functions
which need to be determined. In the language of \eq\nr{rho_k}, 
two spectral functions play a role: one in which we replace
$\eta\theta \to \phi\phi$,  
$\theta^\dagger\eta^\dagger \to \phi^\dagger\phi^\dagger$; 
another
in which 
$\eta\theta \to \chi\chi$,  
$\theta^\dagger\eta^\dagger \to \chi^\dagger\chi^\dagger$. 
Furthermore, 
as suggested by \fig\ref{fig:exchange}(a), the Schr\"odinger
equation determining the spectral functions  
induces a mixing between the two channels. 

In order to determine the mixing, we consider a quantum-mechanical 
problem with the interaction Hamiltonian in \eq\nr{Hint2}. 
Let us define the Wightman function
\be
 C^{ }_{>}(t) 
 \; \equiv \; 
 \tr \Bigl\{ \hat{\rho} 
 \Bigl[
  \haat{\chi}(\vec{r}) 
  \haat{\chi}(\vec{0}) \, \UI(t;0) \, 
  \haat{\phi}^\dagger(\vec{r})
  \haat{\phi}^\dagger(\vec{0})
 \Bigr] \Bigr\} 
 \;,  \la{W2}
\ee
which corresponds to ``half'' of the process in \fig\ref{fig:exchange}(a). 
Here $\UI$ is the time evolution operator in 
the interaction picture. 
The density matrix $\hat{\rho}$ is assumed to have the
form $\hat{\rho} \equiv \mathcal{Z}_0^{-1} e^{-\haat{H}_0/T}
\otimes |0\rangle\langle 0 |$, where $\haat{H}_0$ is 
the Hamiltonian of the Standard Model and $|0\rangle$
is the vacuum state in the sector of the Hilbert space
containing the dark particles. The time evolution operator
can be expanded as usual, 
\be
 \UI(t;0) = 
 \mathbbm{1} - i \int_0^t \! {\rm d}t_1 \, \haat{H}_\rmi{int}(t_1) 
 - \int_0^t \! {\rm d}t_1 \int_0^{t_1} \! {\rm d}t_2
 \, \haat{H}_\rmi{int}(t_1) \, \haat{H}_\rmi{int}(t_2)
 + \rmO(g^3)
 \;. \la{UI}
\ee
The heavy particles can be dealt with 
by making use of canonical commutation relations,  
$[\haat{\phi}(\vec{x}),\haat{\phi}^\dagger(\vec{y})] 
= \delta^{(3)}(\vec{x-y})$, 
etc. 
Thereby a non-zero contraction is obtained which contains the gauge
correlator
\be
 \mathcal{V}^{ }_{\chi\phi}(t) \; \equiv \;
 - g^2 
 \int_0^t \! {\rm d}t_1 \int_0^{t_1} \! {\rm d}t_2 \, 
 e^{i \Delta M(t_1 + t_2)}
 \Bigl\langle
  \haat{Z}^{ }_0(t^{ }_1,\vec{r}) 
  \haat{Z}^{ }_0(t^{ }_2,\vec{0}) +
  \haat{Z}^{ }_0(t^{ }_1,\vec{0}) 
  \haat{Z}^{ }_0(t^{ }_2,\vec{r}) 
 \Bigr\rangle
 \;, \la{V}
\ee
where $\langle ... \rangle$ denotes a thermal average with the 
density matrix $\mathcal{Z}_0^{-1} e^{-\haat{H}_0/T}$. 
We can symmetrize the integrand in $t^{ }_1\leftrightarrow t^{ }_2$
by introducing a time-ordered correlator, 
$\langle ... \rangle^{ }_\rmii{T}$. Furthermore, assuming 
parity symmetry, the $Z_0$ propagator can be written  
as an inverse Fourier transform, 
\be
 \fr12 \Bigl\langle
  \haat{Z}^{ }_0(t^{ }_1,\vec{r}) 
  \haat{Z}^{ }_0(t^{ }_2,\vec{0}) +
  \haat{Z}^{ }_0(t^{ }_1,\vec{0}) 
  \haat{Z}^{ }_0(t^{ }_2,\vec{r}) 
 \Bigr\rangle^{ }_\rmii{T}
 = 
 \int_{\omega,\vec{k}} e^{-i\omega(t^{ }_1 - t^{ }_2)
  + i \vec{k}\cdot\vec{r}} \langle Z^{ }_0 Z^{ }_0 
 \rangle^{ }_\rmii{T}(\omega,k)
 \;.  
\ee
Subsequently the time integrals can be carried out: 
\be
 \Phi(t) 
 \; \equiv \;
  \int_0^t \! {\rm d}t_1 \int_0^{t} \! {\rm d}t_2 \, 
  e^{i \Delta M(t_1 + t_2) -i\omega(t^{ }_1 - t^{ }_2)}
  = 
  e^{i \Delta M t}\, 
  \frac{2\sin\bigl[ \frac{(\omega+\Delta M)t}{2}\bigr]}{\omega + \Delta M}
  \frac{2\sin\bigl[ \frac{(\omega-\Delta M)t}{2}\bigr]}{\omega - \Delta M}
 \;. \la{zero}
\ee
Recalling that 
$
 \lim_{t \to\infty} \frac{\sin(x t)}{x} = \pi\delta(x) 
$,
we see that \eq\nr{zero} is proportional to 
$
 \delta(\omega + \Delta M)\delta(\omega - \Delta M)
$
for $t\to\infty$. It thus
yields a vanishing contribution if $\Delta M > 0$.  
This is a reflection of the fact that with non-degenerate masses
and strictly static on-shell states the process in \fig\ref{fig:exchange}(a) 
violates energy conservation. 

Of course, the heavy particles are not exactly 
static, but can move (this is illustrated in \fig\ref{fig:exchange}(a)). 
This permits for the exchange contribution to 
play a role. A way to determine its magnitude is to think of $\Delta M$
as a low-energy parameter, and to view the computation above as a high-energy
matching step. The matching computation can most simply be carried out 
in the limit $\Delta M \to 0$, whereby \eq\nr{zero} becomes
\be
 \lim_{\Delta M \to 0} \Phi(t) = 
 \frac{4\sin^2\bigl(\frac{\omega t}{2} \bigr)}{\omega^2}
 \;. 
\ee
Now we obtain a non-vanishing 
distribution  in the large-$t$ 
limit,  
\be
 \lim_{t\to\infty} i \partial_t 
 \Bigl\{ \lim_{\Delta M \to 0} \Phi(t) \Bigr\} = 
 \lim_{t\to\infty} \frac{2 i  \sin(\omega t)}{\omega} = 
 2\pi i \delta(\omega)
 \;. \la{stat_V}
\ee
Therefore 
the potential from \eq\nr{V} carries non-zero energy, 
\be
  \lim_{t\to\infty} i \partial_t 
  \Bigl\{ \lim_{\Delta M \to 0} \mathcal{V}^{ }_{\chi\phi}(t) \Bigr\}
   = - g^2 \int_{\vec{k}} e^{i \vec{k}\cdot \vec{r}}
 \, i
 \langle Z^{ }_0 Z^{ }_0 
 \rangle^{ }_\rmii{T}(0,k) 
  \;. \la{proof}
\ee 
This is like the $\vec{r}$-dependent part of \eq\nr{master},
but now the contribution mixes 
two different channels. 
Such a ``non-diagonal'' potential was included, 
e.g.,\ in the analysis of ref.~\cite{idm}. 

A similar computation yields also self-energy contributions
($\chi\to\phi$ in \eq\nr{W2}), 
originating from 
the ``crossed terms'' in two appearances of $\haat{H}_\rmi{int}$
in \eq\nr{UI}: 
\be
  \lim_{t\to\infty} i \partial_t 
  \Bigl\{ \lim_{\Delta M \to 0} \mathcal{V}^{ }_{\phi\phi}(t) \Bigr\}
   = g^2 \int_{\vec{k}} 
 \, i
 \langle Z^{ }_0 Z^{ }_0 
 \rangle^{ }_\rmii{T}(0,k) 
  \;. \la{selfE}
\ee 
This is the $\vec{r}$-independent part of \eq\nr{master}.
Recalling the vacuum counterterm $\delta V$ and the form
of the propagator in \eq\nr{Delta_T_limit} 
and 
noting that $\int_{\vec{k}} ( \frac{1}{k^2 + m_\rmii{th}^2} - \frac{1}{k^2} ) 
= -m_\rmi{th} / (4\pi)$,
the real part of \eq\nr{selfE}
amounts to the Salpeter correction discussed around \eq\nr{Salpeter}

Finally, it is amusing to consider the process shown in 
\fig\ref{fig:exchange}(b), which plays a role in the annihilations
shown in \figs\ref{fig:decays}(e) and (f). 
The relevant Wightman function now reads
\be
 D^{ }_{>}(t) 
 \; \equiv \; 
 \tr \Bigl\{ \hat{\rho} 
 \Bigl[
  \haat{\phi}(\vec{0})
  \haat{\chi}(\vec{r}) 
  \, \UI(t;0) \, 
  \haat{\phi}^\dagger(\vec{r})
  \haat{\chi}^\dagger(\vec{0})
 \Bigr] \Bigr\} 
 \;.  \la{W3}
\ee
A non-zero contribution originates from the crossed terms 
in the product $\haat{H}_\rmi{int}(t^{ }_1)\, \haat{H}_\rmi{int}(t^{ }_2)$, 
cf.\ \eq\nr{Hint2}. For $\Delta M \neq 0$ the time dependence
is different from that in \eq\nr{zero}, however for 
$\Delta M\to 0$ it is the same and 
a potential emerges like in \eq\nr{proof}. 
The overall sign is positive, representing a repulsive interaction
in this channel, suppressing such annihilations. 

%%%%%%%%%%%%%%%%%%%%%%%%%%% SUBSECTION %%%%%%%%%%%%%%%%%%%%%%%%%%%%%%%%%%%
%
\subsection{Numerical results}

%%%%%%%%%%%%%%%%%%%%%%%%%%%%%%%%% FIGURE %%%%%%%%%%%%%%%%%%%%%%%%%%%%%%%%%
\begin{figure}[t]

\hspace*{-0.1cm}
\centerline{%
 \epsfxsize=7.5cm\epsfbox{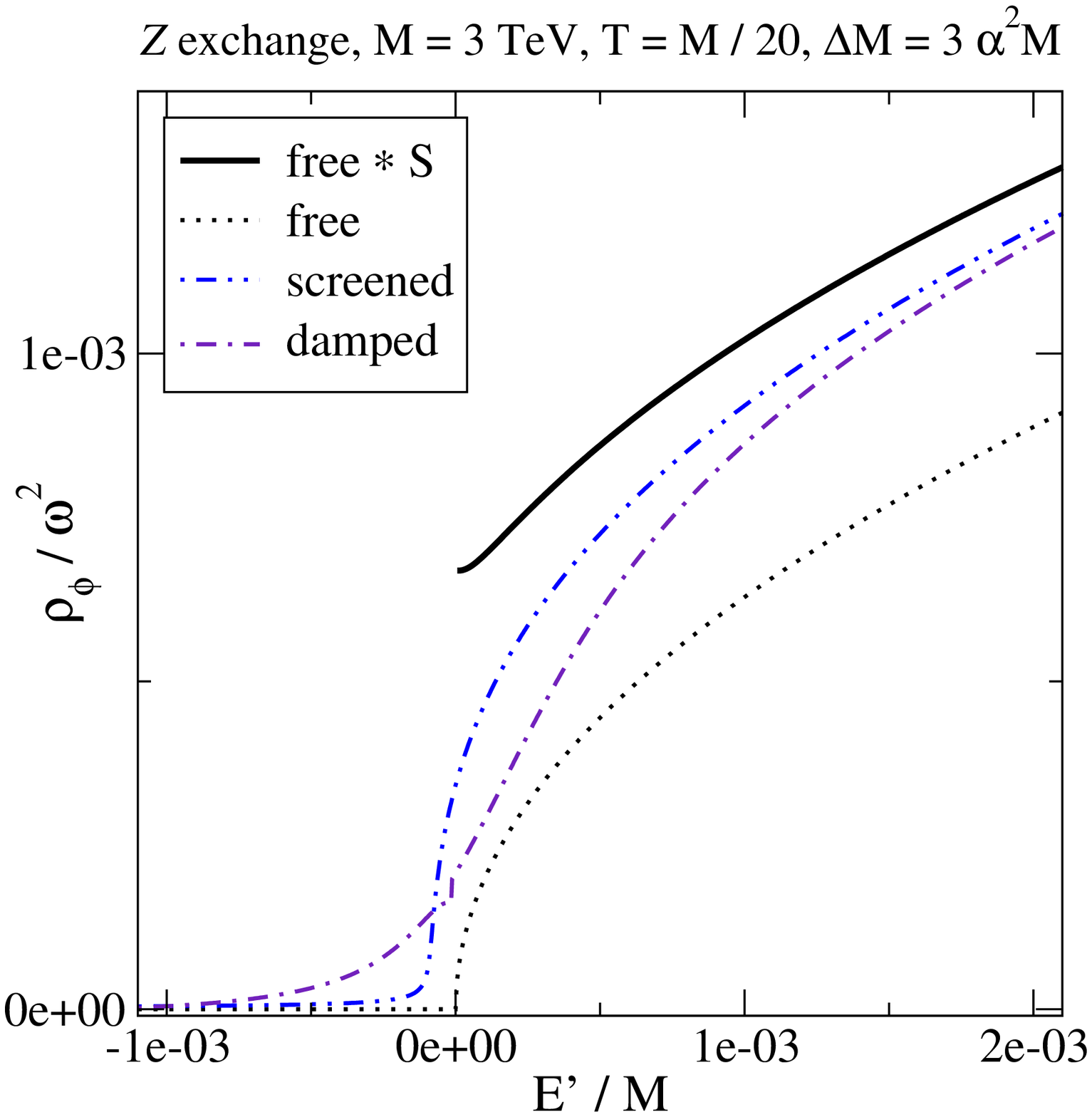}%
 \hspace{0.1cm}
 \epsfxsize=7.5cm\epsfbox{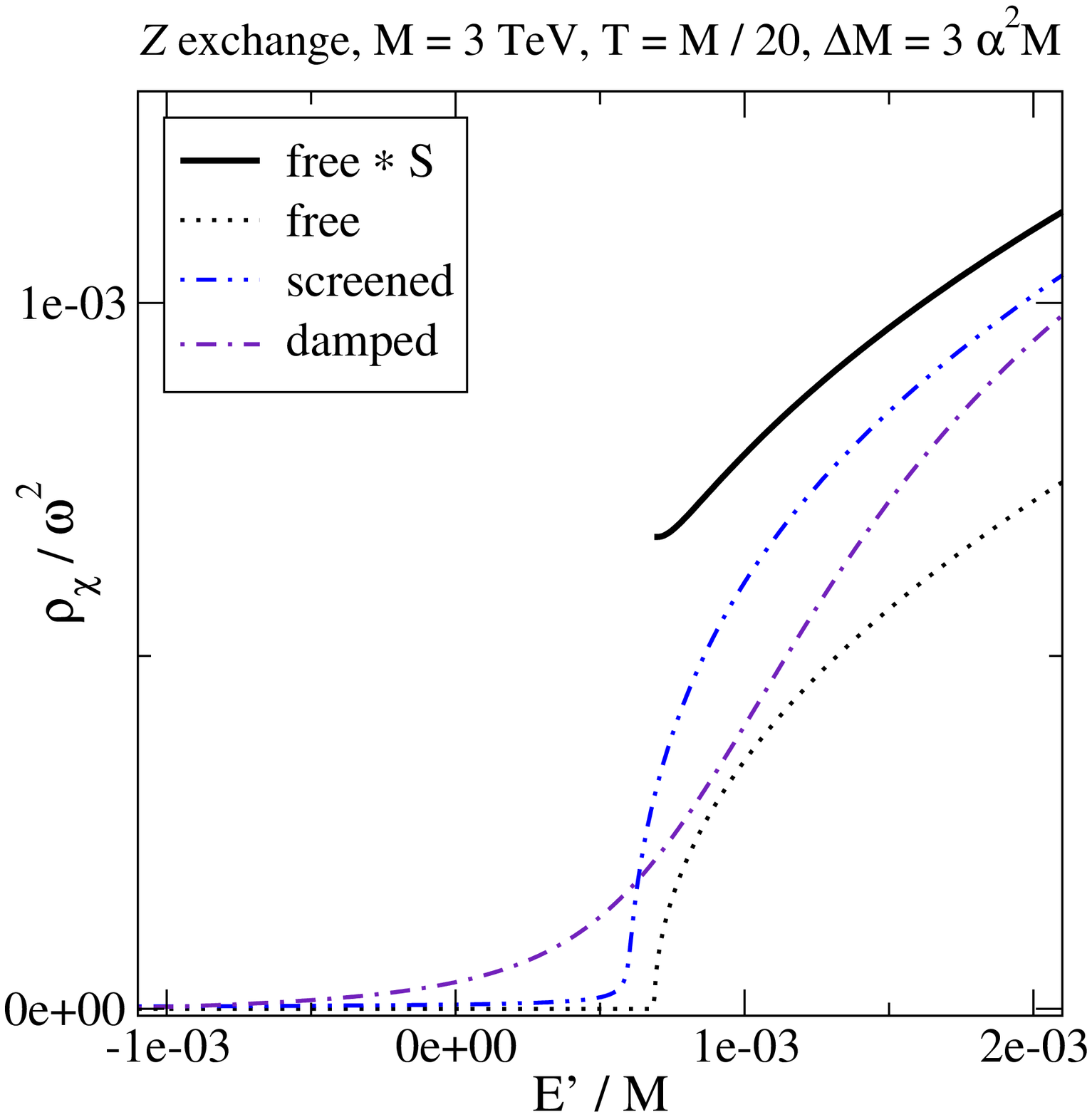}
}

\caption[a]{\small
 Spectral functions obtained from \eqs\nr{Seq_matrix} and \nr{rho_matrix}.
 Left: $\rho^{ }_\phi$. Right: $\rho^{ }_\chi$. 
 By ``screened'' we denote thermal $V$'s 
 but omitting $\Gamma$'s, and by ``damped'' the full
 system. In the former case, 
 small imaginary parts $\Gamma^{ }_{\phi\phi}, 
 \Gamma^{ }_{\chi\chi} \to 0.1 \alpha^2 M \sim 10^{-5} M $ 
 were kept in order to define the spectral function.
}

\la{fig:nondeg}
\end{figure}
%%%%%%%%%%%%%%%%%%%%%%%%%%%%%%%%%%%%%%%%%%%%%%%%%%%%%%%%%%%%%%%%%%%%%%%%%%%

In the presence of the mixing from \eq\nr{proof} 
and assuming $\Delta M \ll M$, 
the equations to be solved amount to a matrix 
version of \eq\nr{Seq}, 
\be
 \biggl( 
  \begin{array}{cc}
  {H}^{ }_{\phi\phi} - i {\Gamma}^{ }_{\phi\phi} - E' & 
  {V}^{ }_{\phi\chi} - i {\Gamma}^{ }_{\phi\chi} \\[1mm]
  {V}^{ }_{\chi\phi} - i {\Gamma}^{ }_{\chi\phi} & 
  2 \Delta{M} + 
  {H}^{ }_{\chi\chi} - i {\Gamma}^{ }_{\chi\chi} - E' \\
  \end{array}
 \biggr)
 \, \biggl( 
  \begin{array}{c}
    G_{\phi} \\ 
    G_{\chi} 
  \end{array}
 \biggr)
  = 
  \biggl( 
  \begin{array}{c}
    \delta^{(3)}(\vec{r-r'}) \\ 
    \delta^{(3)}(\vec{r-r'}) 
  \end{array}
  \biggr)
 \;, \la{Seq_matrix}
\ee
where  
$H^{ }_{\phi\phi} = -\nabla^2/M + V^{ }_{\phi\phi}$; 
$V^{ }_{\phi\phi} = V^{ }_{\chi\chi}$ 
is the $r$-independent part of \eq\nr{Vr};   
$V^{ }_{\phi\chi} = V^{ }_{\chi\phi}$ 
is the $r$-dependent part of \eq\nr{Vr}; 
$\Gamma^{ }_{\phi\phi} = \Gamma^{ }_{\chi\chi}$ 
is $\Gamma(\infty)$ from \eq\nr{Gammar}; 
and $\Gamma^{ }_{\phi\chi} = \Gamma^{ }_{\chi\phi}$ 
is $\Gamma(r) - \Gamma(\infty)$ from \eq\nr{Gammar}. 
Two separate spectral functions are obtained from \eq\nr{rho_rescaled},
\be
 \rho^{ }_{\phi} \;\equiv\; 
 \frac{\alpha M^2}{4\pi}
 \int_0^\infty \! {\rm d}\rho \, \im \biggl[
  \frac{1}{(u^{\phi}_0)^2} \biggr]
 \;, \quad 
 \rho^{ }_\chi \;\equiv\;
 \frac{\alpha M^2}{4\pi}
 \int_0^\infty \! {\rm d}\rho \, \im \biggl[
  \frac{1}{(u^{\chi}_0)^2} \biggr]
 \;. \la{rho_matrix}
\ee
In a vacuum limit these correspond to 
$
  \sum_m |\psi^{ }_m(\vec{0};\pm)|^2 \pi\, \delta(E^{ }_m - E')
$, 
respectively, where $\pm$ denote the upper and lower ``spin'' components
($\phi$ and $\chi$). 

The physics of this system is  
subtle for $E' \lsim 2\Delta M$. In this regime, the $\chi$-pairs
can only appear as ``virtual'' particles. One can imagine that they
are ``integrated out''; it can be shown that
this generates an attractive potential for the $\phi$-pair. 
% The closer we are to the threshold ($E' \to 2\Delta M$), the stronger 
% the attraction, and correspondingly the more complicated the spectrum. 
% The results are also very sensitive to the 
% precise form of $V^{ }_{\chi\phi}$. Fortunately the presence of Debye
% screening simplifies the situation. 
Considering $\Delta {M} = 3 \alpha^2 M$ as an example,
we have solved the equations for two cases:  
the correct Debye-screened potentials
but no widths (``screened''), and the full system including
the widths (``damped''). The results are 
illustrated in \fig\ref{fig:nondeg}.
The screened Sommerfeld enhancement is observed to be 
active even below the second threshold. The shifts
of both thresholds to the left of the ``free'' ones reflect 
the Salpeter correction discussed in \eq\nr{Salpeter}. The inclusion of 
damping smoothens the spectral functions.  After 
integration over the energies according to \eq\nr{Laplace2}, 
thermal effects get however largely hidden, apart from 
an overall suppression by $\exp(-2\Delta M/T)$ of 
annihilations in the $\chi\chi$-channel.  
In a complete phenomenological analysis it should also be 
noted that the $\chi$ particles decay into the $\phi$ ones
after thermal freeze-out, cf.\ e.g.\ ref.~\cite{new2}. 

If $\Delta M$ is increased so that $\Delta M \gg \alpha^2 M$, 
the numerical determination of $\rho^{ }_\phi$ 
becomes challenging,\footnote{%
 The numerics can be modestly 
 accelerated by noting that the off-diagonal terms
 in \eq\nr{Seq_matrix} become small for large distances. Therefore, 
 for $\rho \gg 1$ the equation
 satisfied by the homogeneous solution reads 
 $
  (\partial_\rho^2 + x + i \epsilon) u^{\phi}_0 = 0
 $, 
 and correspondingly for $u^{\chi}_0$. This can be solved as 
 $
 u^{\phi}_0 = C \sin(\rho\sqrt{x+i\epsilon} + \delta)
 $, 
 $C,\delta \in \mathbbm{C}$.  
 Given that
 $
  \sqrt{x+i\epsilon}/\sin^2(\rho\sqrt{x+i\epsilon} + \delta) = 
  - \partial^{ }_{\rho}\cot(\rho\sqrt{x+i\epsilon} + \delta) 
 $,  
 the integral 
 $
  \int_{\rho^{ }_0}^\infty \! {\rm d}\rho \, /(u^{\phi}_0)^2
 $ 
 can be carried out. Both ends contribute, with 
 $
  \lim_{\rho\to\infty} \cot(\rho\sqrt{x+i\epsilon} + \delta) = -i 
 $. 
 Subsequently $C$ and $\delta$ can be traded for 
 $u^{\phi}_0(\rho^{ }_0)$ and 
 $u^{\phi}_0{}'(\rho^{ }_0)$. 
 For $\rho^{ }_0\gg 1$, we thus obtain
 \be
  \int_{\rho^{ }_0}^\infty \! {\rm d}\rho \, 
  \im \biggl[ \frac{1}{(u^{\phi}_0)^2} \biggr] 
  = 
  \im \biggl\{ 
  \frac{1}{u^{\phi}_0(\rho^{ }_0)
  \bigl[ 
   u^{\phi}_0{}'(\rho^{ }_0) - 
   i \sqrt{x+i\epsilon}\, u^{\phi}_0(\rho^{ }_0)
  \bigr]}
  \biggr\} 
  \;. 
 \ee
 In the free limit this result can also be used
 at $\rho^{ }_0 \ll 1$ where, recalling the asymptotics 
 $ u^{\phi}_0(\rho^{ }_0) \approx \rho^{ }_0$, 
 $ u^{\phi}_0{}'(\rho^{ }_0) \approx 1$, 
 it produces
 $\rho^{ }_\phi = \frac{\alpha M^2}{4\pi} \re \sqrt{x + i \epsilon}$, 
 in accordance with \eq\nr{rho_free}.
} 
and a description of the system through a potential model eventually
breaks down. Physically we expect the potential generated
by the virtual exchange to become suppressed for  
$2 \Delta M - E' \gg \alpha^2 M$, 
and correspondingly
the Sommerfeld enhancement experienced by the $\phi$-particles to only
be re-instated somewhat below the heavier threshold, but it would be 
interesting to understand this quantitatively.  

%%%%%%%%%%%%%%%%%%%%%%%%%%% SUBSECTION %%%%%%%%%%%%%%%%%%%%%%%%%%%%%%%%%%%
%
\subsection{Summary of the non-degenerate situation}

The purpose of this section has been to show that
Sommerfeld enhancement does
remain active when $\Delta M \sim \alpha^2 M > 0$. To be more precise, 
there are different cases of gauge exchange between 
non-degenerate particles. With the process in \fig\ref{fig:exchange}(b),
it is possible to have a kinematically permitted 
configuration with static on-shell DM and DM$'$ particles and 
the energy flow $\pm \Delta M$ through the gauge line. 
Therefore the nature of gauge exchange gets modified
only if $\Delta M  \gsim m_\rmi{th}\sim \alpha^{1/2} T$. 
In contrast, the process in \fig\ref{fig:exchange}(a)
leads to a non-trivial 
quantum-mechanical behaviour. Naively, one could think that 
if we are below the threshold for the production
of the heavier particles ($E^{ }_\rmi{kin} < 2 \Delta M$), 
the lighter ones have no partners
to interact with, and they should feel no Sommerfeld enhancement. 
This is not true: the heavier ones can appear as virtual 
states, and in fact they thereby generate an attractive interaction 
between the lighter ones. Therefore, at least if 
$\Delta M \lsim \alpha^2 M$, 
the Sommerfeld effect is present even below the heavier threshold, 
just suppressed somewhat by Debye screening. 

%%%%%%%%%%%%%%%%%%%%%%%%%%% SECTION %%%%%%%%%%%%%%%%%%%%%%%%%%%%%%%%%%%%%%
%
\section{Effects from different colour and spin decompositions}
\la{se:decomp}

If the gauge group is unbroken and non-Abelian, 
then the annihilating pair can appear
in different (global) gauge representations. Within perturbation 
theory the representation dictates whether the gauge force between
the two particles is attractive or repulsive. Presumably, the different
representations appear with specific weights in the (perturbative)
thermal ensemble. 
Thereby the total annihilation rate is a certain combination
of the contributions of the different gauge representations
(for a discussion cf.~e.g.~ref.~\cite{thermal}). The purpose of this section
is to recall how the contributions of all gauge decompositions, 
and also of the various spin states, 
can be included with their proper thermal weights and in a 
gauge-independent manner in the total thermal annihilation rate. 

Within the NRQCD framework, 
annihilations through various gauge and spin channels, 
as well as channels suppressed by higher powers of the relative 
velocity, correspond to unique local gauge-invariant four-particle 
operators~\cite{bodwin}. The four-particle
operators originate from integrating out the energy scale 
$2 M \gg \pi T$; therefore, the determination of the coefficients
can be carried out with vacuum perturbation theory. 

Thermal effects originate when we compute the thermal expectations
values of the operators, in the sense of \eq\nr{def_Gamma}. Assuming
now $\eta$ and $\theta$ to be 2-component spinors, spin effects originate
from structures of the type $\eta^T\! \sigma_i \theta$, and 
gauge effects from the type $\eta^T T^a \theta$, 
where $\sigma_i$ is a Pauli matrix and 
$T^a$ is a generator of the gauge group. 
The sum over $m$ in \eq\nr{def_Gamma} is taken over the full 
ensemble. A spectral function can  
be defined like in \se\ref{ss:rho}, and the total rate from 
every particular operator
reduced to its Laplace transform like in \eq\nr{Laplace2}. 

The essential question is how the Schr\"odinger equation of \se\ref{ss:Sch}
depends on the channel in question. The source term in \eq\nr{Seq}, 
which is independent of the coupling, is modified in a trivial 
way, with $N$ replaced by an appropriate factor. 
The dynamical information concerning the attractive
or repulsive nature of the interaction is encoded in the potential
$V$ and the width $\Gamma$, to be computed in the appropriate
representation.\footnote{%  
 It should be noted that in non-singlet channels 
 the spectral function is {\em not} manifestly gauge independent; 
 nevertheless the total annihilation rate, which can also 
 be measured non-perturbatively, is so~\cite{4quark_lattice}.}

It may be asked whether the Schr\"odinger equations for the 
different channels couple to each other, similarly
to \eq\nr{Seq_matrix}. 
In general, different gauge representations do {\em not} couple. 
In order to illustrate the argument in concrete terms, 
consider the QCD-like decomposition 
$\mathbf{3}\otimes \mathbf{3^*} = \mathbf{1} \oplus \mathbf{8}$.
The symmetry in question is, however, a gauge symmetry: a singlet
representation can convert into an octet only by simultaneously
emitting a colour-electric dipole $\sim \vec{r}\cdot g\vec{E}^a$, or 
another excitation with the same quantum numbers. Since these are 
not among our effective low-energy variables, a mixing is forbidden. 
Indeed, within the PNRQCD framework,  
the width $\Gamma$ appearing in the singlet channel can be shown
to get a contribution precisely from the possibility that the 
singlet split into an octet and a colour-electric-dipole, 
after integrating out the latter two~\cite{jacopo}. Therefore
the octet states have already been accounted for within
the singlet computation.  

For the case of spin channels, we also 
expect orthogonality in general, 
given that gauge exchange is spin independent
to leading order in $1/M$. At higher orders, 
the presence or not of a coupling can be checked 
through an analysis like in \se\ref{ss:derivation}, which also 
establishes whether the exchange in the given 
channel is attractive or repulsive. 

To summarize, the first step is to determine all absorptive operators
in the sense of ref.~\cite{bodwin}. In a resummed perturbative approach,
we subsequently compute the spectral functions for each of them, 
and then take the Laplace transform in \eq\nr{Laplace2}. The total
annihilation rate is the sum of the contributions of the various
operators, i.e.\ the channels are summed together at the level of
total rates. 

%%%%%%%%%%%%%%%%%%%%%%%%%%% SECTION %%%%%%%%%%%%%%%%%%%%%%%%%%%%%%%%%%%%%%
%
\section{Conclusions}
\la{se:concl}

The purpose of this paper has been to revisit the
$s$-wave thermal annihilation rate of massive neutral
particles relevant for cosmology. The formalism is based on
non-relativistic effective theories~\cite{nrqcd,bodwin} in 
combination with a Hard Thermal Loop~\cite{ht1,ht2,ht3,ht4} 
resummed treatment of thermal contributions. The basic object 
addressed is a spectral function, 
i.e.\ the imaginary part of a Green's function, which can be 
interpreted as a differential annihilation rate.  
The total annihilation rate is obtained from
a Laplace transform of the spectral function, 
cf.\ \eq\nr{Laplace2}. The dark matter particles are assumed to interact
through a ``mediator'', which for illustration is taken to be 
a gauge field, characterized by a fine-structure constant $\alpha$.

The Laplace transform in \eq\nr{Laplace2} shows that the spectral
function is needed for $|\omega - 2 M| \,\lsim\, \pi T \ll M$, 
i.e.\ deep in the non-relativistic regime. 
Even though NLO computations of thermal corrections, and higher-order
computations of vacuum corrections, have been carried out for spectral
functions of this type, and even though they do yield formally 
well-behaved results, they show in general poor convergence. Moreover,
a strict NLO computation suggests 
that thermal corrections are power-suppressed
(cf.\ e.g.\ refs.~\cite{dhr,tw}), 
which is not the case in general (cf.\ the Salpeter 
correction in \eq\nr{Salpeter}). To properly understand 
the system in the non-relativistic regime 
therefore requires a resummed treatment. 

Resummations can be implemented through a numerical solution of an 
inhomogeneous Schr\"odinger equation 
(cf.\ \eqs\nr{Seq} and \nr{get_rho}), 
with a static potential incorporating thermal corrections such as 
Debye screening and Landau damping. The latter originates from  
real scatterings of the mediators with plasma particles, as 
is illustrated in some detail around \eqs\nr{imag1} and \nr{imag2}.  
Our hope is that
theoretical uncertainties of freeze-out computations can be
scrutinized and ultimately reduced through this approach. 

In terms of power counting, thermal effects on the 
differential annihilation rate around the threshold 
($|\omega - 2M| \lsim \alpha^2 M$) 
are of order unity for $T \gsim \alpha M$ (cf.\ \eq\nr{TvsM}). In contrast, 
the total annihilation rate gets an $\gsim \rmO(1)$ contribution from 
the threshold region only for $T \lsim \alpha^2 M$ (cf.\ \eq\nr{TvsM2}). 
For weak interactions with $\alpha \sim 0.01$, the freeze-out regime
$T\sim M/25 ... M/20$
corresponds roughly speaking to $T \gsim \alpha M$. Therefore we 
expect a large thermal effect on the differential annihilation 
rate but only a higher-order correction to the total rate. 
For strong interactions with 
$\alpha = g^2 \CF / (4\pi) \gsim 0.1$, in contrast, the 
freeze-out regime may correspond to $T \sim \alpha^2 M$, and  
the threshold region could dominate the total rate.   

In order to consolidate these parametric estimates, we have carried
out numerical studies of semi-realistic models. 
For a purely weakly interacting case 
along the classic WIMP paradigm, 
our basic finding is that even if bound
states were to exist at zero temperature, they are completely melted 
around the freeze-out temperature
(cf.\ \figs\ref{fig:Z0}, \ref{fig:Z0p}). The spectral function does get
smoothened across the two-particle threshold by thermal effects.
Nevertheless, for TeV range masses,  
the total annihilation rate, which gets a contribution
from a broad energy range, is remarkably well (within $\sim 1$\%) 
predicted by a thermally averaged purely Coulombic Sommerfeld factor, 
and even better if Debye screening is accounted for. 

Permitting for some non-degeneracy in the dark particle spectrum, 
we subsequently demonstrated
that the details of the ``coupled-channel'' dynamics 
are delicate (cf.\ \se\ref{se:nondeg}). 
If the mass splitting is not too large, 
we however expect the Sommerfeld enhancement, modified by thermal screening, 
to remain active even below the heavier threshold
(cf.\ \fig\ref{fig:nondeg}).  

Apart from weakly interacting cases, there are models
involving strongly interacting dark matter candidates, or 
strongly interacting particles interacting with the dark matter ones.
In this paper we considered the case of gluinos, for which  
the importance of bound-state effects had already been 
recognized and treated through a phenomenological 
modification of Boltzmann equations~\cite{old5,old51}. We confirm 
that bound states persist up to the temperatures relevant for the 
freeze-out process (cf.\ \fig\ref{fig:gluon}), 
and can boost the annihilation
rate by a factor $\sim 4...80$ compared with a Sommerfeld-enhanced
computation which in itself boosts the annihilation
rate by a similar factor compared with a naive estimate
(cf.\ \fig\ref{fig:scan}). The numerically coincident
magnitude of the two effects is in nice accordance with the parametric
estimate around \eq\nr{TvsM2}, showing that both effects become
large in the same temperature range.  
We stress that within our formalism the existence or melting of bound 
states does not need to be known in advance, but comes out 
from the analysis. Evaluating the phenomenological significance
of these findings requires a complete model-specific study,
which goes beyond the scope of this paper. 

%%%%%%%%%%%%%%%%%%%%%%%%%%% SECTION %%%%%%%%%%%%%%%%%%%%%%%%%%%%%%%%%%
%
\section*{Acknowledgements}

M.L.\ thanks S.~Biondini and M.~Garny for valuable discussions. 
S.K.\ was supported by the
National Research Foundation of Korea under
grant No.\ 2015R1A2A2A01005916
funded by the Korean government (MEST).
M.L.\ was supported by the Swiss National Science Foundation
(SNF) under grant 200020-168988, 
and by the Munich Institute for Astro- and Particle
Physics (MIAPP) of the DFG cluster of excellence 
``Origin and Structure of the Universe''.

%%%%%%%%%%%%%%%%%%%%%%% APPENDIX %%%%%%%%%%%%%%%%%%%%%%%%%%%%%%%%%%%
%
\appendix
\renewcommand{\thesection}{Appendix~\Alph{section}}
\renewcommand{\thesubsection}{\Alph{section}.\arabic{subsection}}
\renewcommand{\theequation}{\Alph{section}.\arabic{equation}}

%%%%%%%%%%%%%%%%%%%%%%%%%%% SECTION %%%%%%%%%%%%%%%%%%%%%%%%%%%%%%%%%%%%%%
%
\section{Neutral gauge field self-energies in the Standard Model}

We present in this appendix the 1-loop thermal self-energy matrix of 
neutral gauge bosons in the Standard Model. Results are given in a general
$R^{ }_\xi$ gauge, and amount to simple generalizations of 
classic results for the vacuum case (cf.\ e.g.\ ref.~\cite{wh}
and references therein). Only terms contributing to the transverse part 
of the self-energy are shown. We introduce the notation 
\ba
     A^{ }_{ }(m) & \equiv & 
     \Tint{P} \frac{1}
     {P^2 + m^2}
     \;, \la{A} \\
     B^{ }_{ }(K;m_1^{ },m_2^{ }) & \equiv & 
     \Tint{P\,} \frac{1}
     {(P^2+m_1^2)[(P+K)^2+m_2^2]}
     \;, \hspace*{7mm} \\ 
     B^{ }_{\mu\nu}(K;m_1^{ },m_2^{ }) & \equiv & 
     \Tint{P\,} \frac{P^{ }_\mu P^{ }_\nu}
     {(P^2+m_1^2)[(P+K)^2+m_2^2]}
     \;, \la{Bmn}
\ea
where $K = (k^{ }_n,\vec{k})$ is a Euclidean four-vector  
and the imaginary-time formalism is employed. The sum-integrals 
$\Tinti{P}$ and $\Tinti{\{P\}}$ go over bosonic and fermionic Matsubara
momenta, respectively; in the fermionic case the structures are 
denoted by 
$\widetilde{A}$, 
$\widetilde{B}$ and 
$\widetilde{B}^{ }_{\mu\nu}$. 
With this notation and letting $\mW' \equiv \xi^{1/2}\mW$, where $\xi$
is a gauge parameter, the hypercharge part of 
the (bare) transverse self-energy matrix reads
\ba
 \Pi^\rmii{ }_{11;\mu\nu} & = & g_1^2 \biggl\{ 
  - B^{ }_{\mu\nu}(K^{ };\mh^{ },\mZ^{ })
  - \delta^{ }_{\mu\nu} \mZ^2\, B(K^{ };\mh^{ },\mZ^{ })
  -  \frac{\delta^{ }_{\mu\nu} A(\mh^{ })}{2}
 \nn & - &  
   2 B^{ }_{\mu\nu}(K^{ };\mW^{ },\mW') 
 - 2\, \delta^{ }_{\mu\nu} \mW^2\, B^{ }_{ }(K^{ };\mW^{ },\mW')  
 + B^{ }_{\mu\nu}(K^{ };\mW',\mW') 
 \nn & + & 
 \frac{17}{3} \biggl[ \widetilde{B}^{ }_{\mu\nu}(K;m^{ }_t,m^{ }_t)
 - \frac{\delta^{ }_{\mu\nu} \widetilde{A}(m^{ }_t)}{2}
 + \delta^{ }_{\mu\nu} \Bigl( \frac{K^2}{4} + \frac{9 m_t^2}{34} \Bigr) 
 \widetilde{B}^{ }_{ }(K;m^{ }_t,m^{ }_t) \biggr]
 \nn & + & 
 \frac{40\nG - 17}{3} \biggl[ \widetilde{B}^{ }_{\mu\nu}(K;0,0)
 - \frac{\delta^{ }_{\mu\nu} \widetilde{A}(0)}{2}
 + \frac{\delta^{ }_{\mu\nu} K^2 \widetilde{B}^{ }_{ }(K;0,0) }{4}  
  \biggr]
 \nn & - & 
   \frac{(D-1)\,\delta^{ }_{\mu\nu}}{2\mh^2}
  \Bigl[ 2 \mW^2 A(\mW^{ }) + \mZ^2 A(\mZ^{ }) \Bigr]
 + \frac{6 \delta^{ }_{\mu\nu} m_t^2 \widetilde{A}(m_t) }{\mh^2}
 \biggr\}  
 \;. \la{PiT11}
\ea 
Here $\mh$ and $m^{ }_t$ are the Higgs and top masses, 
$\nG = 3$ is the number of generations, and 
$D = 4 - 2\epsilon$ is the dimensionality of space-time. 
The mixed part takes the form
\ba
 \Pi^\rmii{ }_{12;\mu\nu} & = & g^{ }_1 g^{ }_2 \biggl\{ 
  - B^{ }_{\mu\nu}(K^{ };\mh^{ },\mZ^{ })
  - \delta^{ }_{\mu\nu} \mZ^2\, B(K^{ };\mh^{ },\mZ^{ })
  - \frac{\delta^{ }_{\mu\nu} A(\mh^{ })}{2}
 \nn &  + &
   B^{ }_{\mu\nu}(K^{ };\mW',\mW') 
  - \delta^{ }_{\mu\nu} \, A(\mW')  
 \nn & - & 
  \widetilde{B}^{ }_{\mu\nu}(K;m^{ }_t,m^{ }_t)
 + \frac{\delta^{ }_{\mu\nu} \widetilde{A}(m^{ }_t)}{2}
 + \delta^{ }_{\mu\nu} \Bigl( - \frac{K^2}{4} + \frac{3 m_t^2}{2} \Bigr) 
 \widetilde{B}^{ }_{ }(K;m^{ }_t,m^{ }_t) 
 \nn & + & 
  \widetilde{B}^{ }_{\mu\nu}(K;0,0)
 - \frac{\delta^{ }_{\mu\nu} \widetilde{A}(0)}{2}
 + \frac{\delta^{ }_{\mu\nu} K^2 \widetilde{B}^{ }_{ }(K;0,0) }{4}  
  \nn  & - &
 \frac{(D-1)\,\delta^{ }_{\mu\nu}}{2\mh^2}
  \Bigl[ 2 \mW^2 A(\mW^{ }) + \mZ^2 A(\mZ^{ }) \Bigr]
  + \frac{6 \delta^{ }_{\mu\nu} m_t^2 \widetilde{A}(m_t) }{\mh^2}
 \biggr\}  
 \;. \la{PiT12}
\ea 
Finally, the SU(2) part can be expressed as 
\ba
 \Pi^\rmii{ }_{22;\mu\nu} & = & g_2^2 \biggl\{ 
  - B^{ }_{\mu\nu}(K^{ };\mh^{ },\mZ^{ })
  - \delta^{ }_{\mu\nu} \mZ^2\, B(K^{ };\mh^{ },\mZ^{ })
  - \frac{\delta^{ }_{\mu\nu} A(\mh^{ })}{2} 
 \nn & + &  
   \Bigl(1 - \frac{K^4}{\mW^4} \Bigr) B^{ }_{\mu\nu}(K^{ };\mW',\mW') 
  + 
  2 \Bigl( 1 + \frac{K^2}{\mW^2} \Bigr)^2 B^{ }_{\mu\nu}(K^{ };\mW^{ },\mW') 
 \nn & + & 
   2\, \delta^{ }_{\mu\nu} 
  \frac{(K^2 + \mW^2)^2}{\mW^2}\, 
  \bigl[ B^{ }_{ }(K^{ };\mW^{ },\mW') - B^{ }_{ }(K^{ };\mW^{ },\mW^{ })
  \bigr]
 \nn & - & 
 4 \Bigl(D - 1 + \frac{K^2}{\mW^2} + \frac{K^4}{4\mW^4} \Bigr)
 B^{ }_{\mu\nu}(K^{ };\mW^{ },\mW^{ }) + 
 2 \delta^{ }_{\mu\nu} (\mW^2 - 2 K^2) \, B(K^{ };\mW^{ },\mW^{ })
 \nn & + & 
  2\, \delta^{ }_{\mu\nu} \Bigl( D - 1 + \frac{K^2}{\mW^2} \Bigr)\, A(\mW^{ })
 -  2\, \delta^{ }_{\mu\nu} \Bigl( 1 + \frac{K^2}{\mW^2} \Bigr) A(\mW')
 \nn & + & 
 3\, \biggl[ \widetilde{B}^{ }_{\mu\nu}(K;m^{ }_t,m^{ }_t)
 - \frac{\delta^{ }_{\mu\nu} \widetilde{A}(m^{ }_t)}{2}
 + \delta^{ }_{\mu\nu} \Bigl( \frac{K^2}{4} + \frac{m_t^2}{2} \Bigr) 
 \widetilde{B}^{ }_{ }(K;m^{ }_t,m^{ }_t) \biggr]
 \nn & + & 
 \bigl( 8\nG - 3 \bigr) \, \biggl[ \widetilde{B}^{ }_{\mu\nu}(K;0,0)
 - \frac{\delta^{ }_{\mu\nu} \widetilde{A}(0)}{2}
 + \frac{\delta^{ }_{\mu\nu} K^2 \widetilde{B}^{ }_{ }(K;0,0) }{4}  
  \biggr]
 \nn & - & 
   \frac{(D-1)\,\delta^{ }_{\mu\nu}}{2\mh^2}
  \Bigl[ 2 \mW^2 A(\mW^{ }) + \mZ^2 A(\mZ^{ }) \Bigr]
 + \frac{6 \delta^{ }_{\mu\nu} m_t^2 \widetilde{A}(m_t) }{\mh^2}
 \biggr\} 
 \;.  \la{PiT22}
\ea 

The self-energies in \eqs\nr{PiT11}--\nr{PiT22} are gauge dependent. 
Gauge-independent expressions are obtained for two linear combinations, 
$
 \Pi^\rmii{ }_{\rmii{$Z$}} \equiv
 \sin^2(\theta)\, \Pi^\rmii{ }_{11}
 + \sin(2\theta)\, \Pi^\rmii{ }_{12}
 + \cos^2(\theta)\, \Pi^\rmii{ }_{22}
$
evaluated at the $Z$ pole $K = -i \mZ$, 
and 
$
 \Pi^\rmii{ }_{\gamma} \equiv 
 \cos^2(\theta)\, \Pi^\rmii{ }_{11}
 - \sin(2\theta)\, \Pi^\rmii{ }_{12}
 + \sin^2(\theta)\, \Pi^\rmii{ }_{22}
$
evaluate at the $\gamma$ pole $K = 0$. However at $\pi T \gg \mZ$ the vacuum
poles are no longer relevant. Indeed there is a third limit, 
the so-called  Hard Thermal Loop (HTL) one~\cite{ht1,ht2,ht3,ht4}, 
in which \eqs\nr{PiT11}--\nr{PiT22} are separately gauge-independent, 
as we now show. 

As a first step, let us write down the thermal parts of the ``master''
sum-integrals in \eqs\nr{A}--\nr{Bmn}. For this purpose we keep 
$k\equiv |\vec{k}| \neq 0$ and work out the expressions up to and
including $\rmO(\omega)$ after the analytic continuation 
$k_n\to -i [\omega+ i 0^+]$. Denoting 
$
 \int_{\vec{p}} = \fr12 \int_{-1}^{+1}\! {\rm d}z\, \int_p  
$,
where $z$ is an angular variable, and omitting the vacuum parts, 
we get
\ba
 A^{(T)}(m) & = & \int_{p} \frac{\nB{}(\epsilon^{ })}{\epsilon^{ }}
 \;, \\ 
 & & \hspace*{-2.3cm}
 B^{(T)}(-i[\omega+i0^+],\vec{k};m^{ }_1,m^{} _2) \nn 
 & = & 
 \int_p \biggl\{
 \frac{\nB{}(\epsilon^{ }_1)}{4pk \epsilon^{ }_1}
 \ln \biggl| \frac{m_2^2 - m_1^2 + k^2 + 2 pk}
                  {m_2^2 - m_1^2 + k^2 - 2 pk} \biggr|
 + 
  \frac{\nB{}(\epsilon^{ }_2)}{4pk \epsilon^{ }_2}
 \ln \biggl| \frac{m_1^2 - m_2^2 + k^2 + 2 pk}
                  {m_1^2 - m_2^2 + k^2 - 2 pk} \biggr| \biggr\}
 \nn & + & 
 \frac{i \omega}{8\pi k T}
 \int_{\epsilon^{ }_\rmii{min}}^{\infty} \! {\rm d}\epsilon \, 
 \nB{}(\epsilon) \bigl[ 1 + \nB{}(\epsilon) \bigr]
 + \rmO(\omega^2)
 \;, \\  
 & & \hspace*{-2.3cm}
 B^{(T)}_{00}(-i[\omega+i0^+],\vec{k};m^{ }_1,m^{} _2) \nn
 & = & - 
 \int_p \biggl\{ 
 \frac{\epsilon^{ }_1 \nB{}(\epsilon^{ }_1)}{4pk }
 \ln \biggl| \frac{m_2^2 - m_1^2 + k^2 + 2 pk}
                  {m_2^2 - m_1^2 + k^2 - 2 pk} \biggr|
 +
 \frac{\epsilon^{ }_2 \nB{}(\epsilon^{ }_2)}{4pk }
 \ln \biggl| \frac{m_1^2 - m_2^2 + k^2 + 2 pk}
                  {m_1^2 - m_2^2 + k^2 - 2 pk} \biggr| \biggr\}
 \nn & - & 
 \frac{i \omega}{8\pi k T}
 \int_{\epsilon^{ }_\rmii{min}}^{\infty} \! {\rm d}\epsilon \, \epsilon^2 \,  
 \nB{}(\epsilon) \bigl[ 1 + \nB{}(\epsilon) \bigr]
 + \rmO(\omega^2)
 \;, \hspace*{7mm} \la{B_T_expl}
\ea
where
\be
 \epsilon^{  }_i \; \equiv \; \sqrt{p^2 + m_i^2}
 \;, \quad 
 \epsilon^{ }_\rmi{min} \; \equiv \;
 \frac{\sqrt{k^4 + 2 k^2(m_1^2 + m_2^2) + (m_1^2 - m_2^2)^2}}{2k}
 \;. 
\ee
The fermionic cases are obtained by replacing $\nB{}\to -\nF{}$.

Given that the imaginary parts play an important role in the analysis, 
let us detail their physical origin. Consider a space-like
vector boson, with energy $\omega$ and momentum $k > \omega$, scattering 
on energetic plasma particles. For illustration,
assume the plasma particles
to be bosons and consider the case that they do not
change their identity in the scattering, i.e.\ $m^{ }_1 = m^{ }_2$.  
Incorporating both reactions and inverse reactions, 
i.e.\ a decay and generation of a vector boson
with 4-momentum $(\omega,\vec{k})$,
the scattering rate for a process in which  
the matrix element is proportional to the energy
of a scatterer takes the form
\ba
 & & \loss \quad - \quad \gain \la{imag1} \\[2mm] 
 & = & 
 \int_{\vec{p}} \frac{\epsilon_p^2}{4 \epsilon^{ }_p \epsilon^{ }_{p+k}}
 \Bigl\{ 
  \nB{}(\epsilon^{ }_p) \bigl[ 1 + \nB{}(\epsilon^{ }_{p+k} ) \bigr] - 
  \nB{}(\epsilon^{ }_{p+k}) \bigl[ 1 + \nB{}(\epsilon^{ }_p ) \bigr] 
 \Bigr\} 
 \, 
 2\pi \delta(\omega + \epsilon^{ }_p - \epsilon^{ }_{p+k})
 \nn  
 & = & 
 \int_{\vec{p}} \frac{\epsilon_p^2}{4 \epsilon^{ }_p \epsilon^{ }_{p+k}}
 \Bigl\{ 
  \nB{}(\epsilon^{ }_p)  - 
  \nB{}(\epsilon^{ }_{p} + \omega) 
 \Bigr\} 
 \, 
 2\pi \delta(\omega + \epsilon^{ }_p - \epsilon^{ }_{p+k})
 \nn  
 & = & 
 - \omega \pi
 \int_{\vec{p}} \frac{\epsilon_p^{ }}{2 \epsilon^{ }_{p+k}}
 \, \nB{}'(\epsilon^{ }_p)   
 \, 
 \delta(\epsilon^{ }_p - \epsilon^{ }_{p+k})
 \; + \; \rmO(\omega^2)
 \nn 
 & = & 
 - \omega \pi
 \int_{\vec{p}} \, \epsilon^{ }_p \,
  \nB{}'(\epsilon^{ }_p)   
 \, 
 \delta(\epsilon^{2}_p - \epsilon^{2}_{p+k})
 \; + \; \rmO(\omega^2)
 \nn  
 & = & 
 - \frac{\omega \pi}{2}
 \int_{p} \, \epsilon^{ }_p \,
  \nB{}'(\epsilon^{ }_p)   
 \,  \int_{-1}^{+1}\! {\rm d}z \, 
 \delta(k^2 + 2 p k z)
 \; + \; \rmO(\omega^2)
 \nn  
 & = & 
 \frac{\omega \pi}{4 k T}
 \int_{p} \, \frac{ \epsilon^{ }_p 
  \nB{}(\epsilon^{ }_p)   
  \bigl[1 +  \nB{}(\epsilon^{ }_p) \bigr] 
  \theta(2 p - k)  }{p} 
 \; + \; \rmO(\omega^2)
 \nn  
 & = & 
 \frac{\omega}{8\pi k T}
 \int_{\epsilon^{ }_\rmii{min}}^{\infty}
 \! {\rm d}\epsilon \, \epsilon^2 \, 
  \nB{}(\epsilon^{ })   
  \bigl[1 +  \nB{}(\epsilon^{ }) \bigr]   
 \; + \; \rmO(\omega^2)
 \;. \la{imag2}
\ea
This is a special case of the last line of \eq\nr{B_T_expl}, and
indicates that the imaginary part originates from 
real scatterings experienced by space-like gauge fields.  

We now turn to the HTL limit~\cite{ht1,ht2,ht3,ht4}. 
It corresponds to the approximation
$\pi T \gg m, k$, 
and concerns terms which scale as $T^2$. 
The sum-integral $B^{(T)}$ is of 
$
 \rmO(\ln (T/m))
$
and therefore gives no HTL structure. 
The non-vanishing HTL structures read, for $D=4$, 
\ba
 A^{(T)} & \to & \frac{T^2}{12} \;, \la{A_htl} \\ 
 B^{(T)}_{00} & \to & -\frac{T^2}{24}\, 
 \biggl[ 1 + \frac{i \omega \pi}{k} + \rmO(\omega^2) \biggr] \;, \\  
 \widetilde{A}^{(T)} & \to & - \frac{T^2}{24} \;, \\ 
 \widetilde{B}^{(T)}_{00} & \to &  \frac{T^2}{48} \, 
 \biggl[ 1 + \frac{i \omega \pi}{k} + \rmO(\omega^2) \biggr] \;. 
 \la{Bmn_htl}
\ea
After inserting these, \eqs\nr{PiT11}--\nr{PiT22} reduce 
to gauge-independent expressions:
\ba
 \Pi^{\rmii{}(T)}_{11;00} & \to &   
 - \frac{g_1^2T^2}{8}
 \biggl( 1 + \frac{2 \mW^2 + \mZ^2 + 2 m_t^2}{\mh^2} \biggr)
% \nn & & \quad
 + \, m_\rmii{E1}^2 % T^2 \biggl( \frac{1}{6} + \frac{5\nG}{9}\biggr)
 \biggl( 1 + \frac{i \omega \pi}{2 k} \biggr)  + \rmO(\omega^2)
 \;, \la{Pi11_htl} \\ 
%%%%%%%%%%%
 \Pi^{\rmii{}(T)}_{12;00} & \to &  
 - \frac{g^{ }_1 g^{ }_2T^2}{8}
  \biggl( 1 + \frac{2 \mW^2 + \mZ^2 + 2 m_t^2}{\mh^2}
 \biggr) + \rmO(\omega^2)
 \;, \\ 
%%%%%%%%%%%
 \Pi^{\rmii{}(T)}_{22;00} & \to &  
 - \frac{g_2^2T^2}{8}
 \biggl( 1 + \frac{2 \mW^2 + \mZ^2 + 2 m_t^2}{\mh^2} \biggr)
% \nn & & \quad
 +\, m_\rmii{E2}^2 % T^2 \biggl( \frac{5}{6} + \frac{\nG}{3}\biggr)
 \biggl( 1 + \frac{i \omega \pi}{2 k} \biggr)  + \rmO(\omega^2)
 \;, \la{Pi22_htl} \hspace*{7mm}
\ea
where $m^2_\rmii{E1}$ and $m^2_\rmii{E2}$ are from \eq\nr{Debye}.  
Combined with tree-level effects from the Higgs mechanism, {\em viz.}
$
 \Pi^{\rmii{}(0)}_{ij;00} = g^{ }_i g^{ }_j \vev^2/4
$,
the first parts of \eqs\nr{Pi11_htl}--\nr{Pi22_htl} can be accounted for
through a $T$-dependent Higgs expectation value, 
$
 \Pi^{\rmii{}(0)}_{ij;00} + 
 \Pi^{\rmii{}(T)}_{ij;00} = g^{ }_i g^{ }_j \vT^2/4 + ...
$,
where
\be
 \vT^2 \; \equiv \; - \frac{m_\phi^2}{\lambda}
 \;\; \mbox{for} \;\; m_\phi^2 < 0  
 \;, 
 \quad
 m_\phi^2 \; \equiv \; - \frac{\mh^2}{2}
 + \frac{( g_1^2 + 3 g_2^2 +8 \lambda  + 4 h_t^2 )T^2}{16}
 \;. \la{v_T}
\ee
Subsequently we redefine $\mW$ and $\mZ$ to stand for the gauge
boson masses defined with $\vT$. 

The propagators corresponding to the HTL self-energies
in \eqs\nr{Pi11_htl}--\nr{Pi22_htl} can be obtained through 
straightforward inversion. Since only the small-$\omega$ limit
is needed, the terms proportional to $\omega$ can be expanded 
to first order. Projecting the matrix subsequently to the 
$Z$ direction, the retarded $Z$ propagator becomes
\be
 \bigl\langle Z^{ }_0 Z^{ }_0 \bigr\rangle^{ }_\rmii{R}  
 = 
 \bigl( \sin\theta \; \cos\theta \bigr)
 \, \Bigl( \prop - \prop\, \Omega\, \prop \Bigr) \, 
 \biggl( 
 \begin{array}{c} 
  \sin\theta \\ \cos\theta
 \end{array}
 \biggr)  \; + \; \rmO(\omega^2) 
 \;, 
\ee
where $\Delta$ can be diagonalized through a rotation
by the angle $\tilde{\theta}$ defined in \eq\nr{mixing},  
\be
 \prop = 
 \Biggl( 
 \begin{array}{rr} 
    \cos\tilde{\theta} & \sin\tilde{\theta}  \\ 
   - \sin\tilde{\theta} & \cos\tilde{\theta}  
 \end{array}  
 \Biggr)
 \, 
 \Biggl( 
 \begin{array}{cc} 
   \frac{1}{k^2 + \mQt^2} & 0 \\ 
   0 & \frac{1}{k^2 + \mZt^2} 
 \end{array}  
 \Biggr)
 \, 
 \Biggl( 
 \begin{array}{rr} 
   \cos\tilde{\theta} & - \sin\tilde{\theta} \\
   \sin\tilde{\theta} & \cos\tilde{\theta} 
 \end{array}  
 \Biggr)
 \;. 
\ee
The masses are given in \eq\nr{mZt_mQt}. 
The width matrix reads
\be
 \Omega = \frac{i\omega\pi}{2k}
 \biggl(
 \begin{array}{cc}
   m_\rmii{E1}^2 & 0 \\ 
   0 & m_\rmii{E2}^2 
 \end{array} 
 \biggr)
 \;.
\ee
Consequently the static limit of the time-ordered propagator, 
\eq\nr{Delta_T_limit},  becomes
\ba
% && \hspace*{-1cm}
 \lim_{\omega\to 0} i 
 \bigl\langle Z^{ }_0 Z^{ }_0 \bigr\rangle^{ }_\rmii{T}(\omega,k)
% \nn 
 & = &   
% \bigl\langle Z^{ }_0 Z^{ }_0 \bigr\rangle^{ }_\rmii{R}(0,k) 
% + i  \lim_{\omega\to 0} 
% \frac{2T 
% }{\omega}
% \im \bigl\langle Z^{ }_0 Z^{ }_0 \bigr\rangle^{ }_\rmii{R}(\omega,k)
% \nn 
% \; = \;  
 \frac{\cos^2(\tilde\theta - \theta)}{k^2 + \mZt^2}  + 
 \frac{\sin^2(\tilde\theta - \theta)}{k^2 + \mQt^2}
 \nn
 & - & \frac{i\pi T}{k}
 \biggl\{ 
  m_\rmii{E1}^2
  \, 
  \biggl[ 
   \frac{\sin\tilde{\theta} \cos(\tilde\theta - \theta)}
        {k^2 + \mZt^2} - 
   \frac{\cos\tilde{\theta} \sin(\tilde\theta - \theta)}
        {k^2 + \mQt^2} 
  \biggr]^2
 \nn &  & 
 \hspace*{6mm} + \; 
  m_\rmii{E2}^2
  \, 
  \biggl[ 
   \frac{\cos\tilde{\theta} \cos(\tilde\theta - \theta)}
        {k^2 + \mZt^2} + 
   \frac{\sin\tilde{\theta} \sin(\tilde\theta - \theta)}
        {k^2 + \mQt^2} 
  \biggr]^2
 \biggr\}
 \;. \hspace*{6mm} \la{Z0_prop}
\ea
The two terms in the imaginary part correspond to real scatterings
through hypercharge and weak interactions, respectively.
Eq.~\nr{Z0_prop} leads directly to \eqs\nr{Vr} and \nr{Gammar}. 

Finally, for completeness, we also give the time-ordered 
$W$ propagator:
\ba
 \lim_{\omega\to 0} i 
 \bigl\langle W^{a }_0 W^{a' }_0 \bigr\rangle^{ }_\rmii{T}(\omega,k)
 & = &   
 \delta^{aa'}
 \biggl\{ 
 \frac{1}{k^2 + \mWt^2}
 \; - \; \frac{i\pi T}{k}
 \frac{m_\rmii{E2}^2}
        {(k^2 + \mWt^2)^2}
 \biggr\}
 \;. \hspace*{6mm} \la{W0_prop}
\ea
Here $a,a'\in\{1,2\}$ and 
$\mWt^2 \equiv \mW^2 + m_\rmii{E2}^2$, 
where $m_\rmii{E2}^2$ is from \eq\nr{Debye}.

%%%%%%%%%%%%%%%%%%%%%%%%%%% SECTION %%%%%%%%%%%%%%%%%%%%%%%%%%%%%%%%%%%%%%
%
\section{Gauge field self-energy in a dark U(1) model}

In this appendix 
we present the 1-loop thermal self-energy matrix of $Z'$ gauge bosons
(mass $\mV$)
interacting with a dark scalar field
(mass $\mS$), which breaks the gauge symmetry.
Only the transverse part of the  
self-energy is shown. Making use of the notation in 
\eqs\nr{A}--\nr{Bmn} it reads
\ba
 \Pi^\rmii{}_{\mu\nu} & = & -2 e'{}^2 \biggl\{ 
  2 B^{ }_{\mu\nu}(K^{ };\mS^{ },\mV^{ })
  + 2 \delta^{ }_{\mu\nu} \mV^2\, B(K^{ };\mS^{ },\mV^{ })
 \nn & + & 
 \delta^{ }_{\mu\nu} A(\mS^{ })
  +  \frac{(D-1)\,\delta^{ }_{\mu\nu}  \mV^2 A(\mV^{ }) 
 }{\mS^2}
 \biggr\}  
 \;. \la{PiV11}
\ea 
Carrying out analytic continuation, $k_n \to -i (\omega+ i 0^+)$,
and taking the HTL limit, cf.\ \eqs\nr{A_htl}--\nr{Bmn_htl}, we get
\ba
 \Pi^{(T)}_{00} & \to & e'{}^2 \biggl\{ 
 - \frac{T^2}{3} \biggl( 1 + \frac{3 \mV^2}{2 \mS^2} \biggr)
 +  \frac{T^2}{3}
 \biggl( 1 + \frac{i \omega \pi}{2 k} \biggr)  + \rmO(\omega^2)
 \biggr\} 
 \;. \la{PiV_htl} 
\ea
Combined with the tree-level effect from the Higgs mechanism, {\em viz.}
$
 \Pi^{(0)}_{00} = e'{}^2 \vev'{}^2
$,
the first part of \eq\nr{PiV_htl} can be accounted for
through a temperature dependent Higgs expectation value, 
$
 \Pi^{(0)}_{00} + 
 \Pi^{(T)}_{00} = e'{}^2 \vT'\!{}^2 + ...
$,
where
\be
 \vT'\!{}^2 \; \equiv \; - \frac{m_{\phi'}^2}{\lambda'} 
 \;\; \mbox{for} \;\; m_{\phi'}^2 < 0 
 \;, \quad
 m_{\phi'}^2 \; \equiv \; - \frac{\mS^2}{2}
 + \frac{( 3 e'{}^2 +4 \lambda' )T^2}{12}
 \;. \la{vp_T}
\ee
Subsequently we redefine $\mV$
as $\mV \equiv e' \vT'$.
The latter term in \eq\nr{PiV_htl} is parametrized by the Debye mass
$m_\rmii{E$'$}^2$
defined in \eq\nr{mDebyep}. 

The resummed time-ordered propagator can now be 
computed from \eq\nr{Delta_T_limit}, yielding
\ba
 \lim_{\omega\to 0} i 
 \bigl\langle V^{ }_0 V^{ }_0 \bigr\rangle^{ }_\rmii{T}(\omega,k) 
% & = &   
% \bigl\langle V^{ }_0 V^{ }_0 \bigr\rangle^{ }_\rmii{R}(0,k) 
% + i  \lim_{\omega\to 0} 
% \frac{2T 
% }{\omega}
% \im \bigl\langle V^{ }_0 V^{ }_0 \bigr\rangle^{ }_\rmii{R}(\omega,k)
% \nn 
 & = &  
 \frac{1}{k^2 + \mV^2 + m_\rmii{E$'$}^2} 
  -  \frac{i\pi T}{k}
  \frac{ 
  m_\rmii{E$'$}^2
  }{(k^2 + \mV^2 + m_\rmii{E$'$}^2 )^2 }
 \;. \la{DeltaTp}
\ea
This directly leads to \eq\nr{Gammarp}. 

%%%%%%%%%%%%%%%%%%%%%%%%%%%%%%% BIBLIOGRAPHY %%%%%%%%%%%%%%%%%%%%%%%%%%%%%%
%

\end{document}